\documentclass[aps,twocolumn,prb,amsmath,amssymb]{revtex4}

\usepackage{graphicx,epsfig}
\usepackage{color}
\usepackage{wasysym}
\usepackage[normalem]{ulem}
\begin{document}
\title{Mixed electrical-heat noise spectrum in a quantum dot}
\author{Paul Eym\'eoud$^{1,2}$}
\author{Adeline Cr\'epieux$^{1}$}
\affiliation{$^1$ Aix Marseille Univ, Universit\'e de Toulon, CNRS, CPT, Marseille, France}
\affiliation{$^2$ Aix Marseille Univ, CNRS, CINAM, Marseille, France}

\begin{abstract}
Using the Keldysh Green function technique, we calculate the finite-frequency correlator between the electrical current and the heat current flowing through a quantum dot connected to reservoirs. At equilibrium, we find that this quantity, called mixed noise, is linked to the thermoelectric ac-conductance by the fluctuation-dissipation theorem. Out-of-equilibrium, we discuss its spectrum and find evidence of the close relationship between the mixed noise and the thermopower. We study the spectral coherence and identify the conditions to have a strong correlation between the electrical and heat currents. The change in the spectral coherence due to the presence of a temperature gradient between the reservoirs is also highlighted.
\end{abstract}

\maketitle


\section{Introduction}

The electrical noise spectrum in quantum systems is accessible experimentally through very sensitive techniques such as spectrum analyzer~\cite{Picciotto1997,Saminadayar1997}, use of a superconductor-insulator-superconductor tunnel junction as an on-chip spectrum analyzers~\cite{Deblock2003} and measurement of photon emission spectrum~\cite{Schull2009}. It exists also a proposal to measure the heat statistics in quantum devices~\cite{Sanchez2012}. Given the fast progress of detection techniques, it is not forbidden to imagine that in the next few years, the measurement of the noise correlator between the electrical current and the heat current (mixed noise) would be possible.

In parallel, calculations of finite-frequency electrical-heat mixed noise are needed for quantum systems. There exist very few works on the zero-frequency mixed noise~\cite{Giazotto2006,Sanchez2013,Battista2014_proc,Crepieux2015,Crepieux2016} and not even one concerning the finite-frequency mixed noise. Theoretically, the studies are limited to the electrical noise spectrum (see~\cite{Blanter2000,Martin2005,Zamoum2016} and references therein), to the energy noise spectrum~\cite{Averin2010,Sergi2011,Agarwalla2015}, to the statistics of the energy current in the presence of time-dependent excitation~\cite{Battista2014,Yu2016}, and to the heat noise spectrum~\cite{Zhan2011}. This is regrettable since it has been shown recently that the zero-frequency mixed noise contains information on the thermoelectric response of the system~\cite{Crepieux2015,Crepieux2016}: it gives the figure of merit in the linear response regime and it is related to the thermoelectric efficiency in the weak transmission regime (Schottky 
regime). At finite-frequency, the mixed noise should bring information on the 
dynamics of the thermoelectric conversion, in particular on the thermoelectric response of time-modulated systems, which is the study of an increasing number of works \cite{Crepieux2011,Goker2012,Goker2013,Boehnke2013,Bagheri2014,Chirla2014,Zhou2015,Dare2016,Ludovico2016,Okada2016}. In this paper, we fill this lacuna by calculating the mixed noise spectrum of a quantum dot (QD) using the Keldysh out-of-equilibrium Green function technique. We focus on the non-symmetrized noise spectrum since this is the quantity which is relevant for quantum systems, due to the fact that the current operators do not commute with each other \cite{Lesovik1997,Gavish2000,Deblock2003}.

The paper is organized as follows: We present the model and give the definition of electrical and heat currents in Sec.~II. The results for the noise spectra are presented in Sec.~III, and discussed in Sec.~IV. We conclude in Sec.~V.

 
\section{Model}

\begin{figure}[h!]
\begin{center}
\includegraphics[width=7cm]{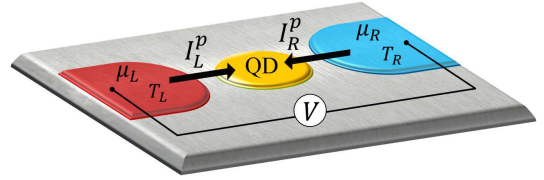}
\caption{Picture of the QD connected to left and right reservoirs with distinct temperatures and chemical potentials (we take $eV=\mu_L-\mu_R$ and $\Delta T=T_L-T_R$). The black arrows indicate the convention chosen for the currents' direction.}
\label{figure1}
\end{center} 
\end{figure}

To model the QD connected to left ($L$) and right ($R$) reservoirs, we use the Hamiltonian $H=H_L+H_R+H_{D}+H_T$,
where $H_{\alpha=L,R}=\sum_{k\in \alpha} \varepsilon_{k \alpha} c_{k \alpha}^{\dag}c_{k \alpha}$ describes the energy of electrons in the reservoir $\alpha$, with $c_{k \alpha}^{\dag}$ ($c_{k \alpha}$), the creation (annihilation) operator, $H_{D}= \varepsilon_d d^{\dag}d$ describes the QD with a single energy level $\varepsilon_d$, with $d^{\dag}$ ($d$) the creation (annihilation) operator, and $H_T=\sum_{\alpha=L,R}\sum_{k\in \alpha}(V_{k \alpha} c_{k \alpha}^{\dag} d+h.c.)$ describes the transfer of electrons from the reservoirs to the QD and vice versa. The left and right reservoirs are assumed to be at equilibrium with temperature $T_{L,R}$ and chemical potential $\mu_{L,R}$ (see Fig.~\ref{figure1}).

The charge current, $\hat{I}^0_{\alpha}$, and heat current, $\hat{I}^1_{\alpha}$, flowing from the reservoir $\alpha$ to the QD, are given by the time derivatives \cite{Jauho1994,Haug2007,Crepieux2012} of the operators \textit{number} of electrons in the reservoir $\alpha$, $N_{\alpha}$, and \textit{energy} of electrons in the reservoir $\alpha$, $H_{\alpha}$: $\hat{I}^0_{\alpha}=-e\dot{N}_{\alpha}$, and  $\hat{I}^1_{\alpha}=-\dot{H}_{\alpha}+\mu_{\alpha}\dot{N}_{\alpha}$,
with $N_{\alpha}=\sum_{k\in\alpha}c^{\dag}_{k \alpha}c_{k \alpha}$. The time derivatives of these two quantities are equal to $\dot{N}_{\alpha}=i\hbar^{-1}\big[H,N_{\alpha}\big]$, and $\dot{H}_{\alpha}=i\hbar^{-1}\big[H,H_{\alpha}\big]$, which lead after calculation to
\begin{eqnarray}\label{N}
\dot{N}_{\alpha}&=&\frac{i}{\hbar}\sum_{k\in\alpha}\big(-V_{k \alpha}c^{\dag}_{k \alpha}d+V^{*}_{k \alpha}d^{\dag}c_{k \alpha}\big)~,\\
\label{H}
\dot{H}_{\alpha}&=&\frac{i}{\hbar}\sum_{k\in\alpha}\varepsilon_{k \alpha}\big(-V_{k \alpha}c^{\dag}_{k \alpha}d+V^{*}_{k \alpha}d^{\dag}c_{k \alpha}\big)~.
\end{eqnarray}

Injecting Eqs.~(\ref{N}) and (\ref{H}) in the definitions of $\hat{I}^0_{\alpha}$ and $\hat{I}^1_{\alpha}$, we obtain: 
\begin{eqnarray*}
\hat{I}^0_{\alpha}&=&\frac{ie}{\hbar}\sum_{k\in\alpha}\big(V_{k \alpha}c^{\dag}_{k \alpha}d-V^{*}_{k \alpha}d^{\dag}c_{k \alpha}\big)~,\\
\hat{I}^1_{\alpha}&=&\frac{i}{\hbar}\sum_{k\in\alpha}(\varepsilon_{k \alpha}-\mu_{\alpha})\big(V_{k \alpha}c^{\dag}_{k \alpha}d-V^{*}_{k \alpha}d^{\dag}c_{k \alpha}\big)~,
\end{eqnarray*}
which give in a compact form
\begin{eqnarray}\label{c1}
\hat I^p_\alpha&=&\frac{ie^{1-p}}{\hbar}\sum_{k\in\alpha} \big(\varepsilon_{k \alpha}-\mu_\alpha\big)^p\big( V_{k \alpha} c_{k \alpha}^{\dag} d-V_{k \alpha}^{\ast} d^{\dag} c_{k \alpha} \big)~.\nonumber\\
\end{eqnarray}


\section{Noise spectrum}

The non-symmetrized noise spectrum is defined as 
\begin{eqnarray}\label{def_S}
\mathcal{S}^{pq}_{\alpha\beta}(\omega)=\int_{-\infty}^{\infty} \mathcal{S}^{pq}_{\alpha\beta}(t,0)e^{-i\omega t}dt~,
\end{eqnarray}
where $\mathcal{S}^{pq}_{\alpha\beta}(t,0)=\langle \Delta \hat{I}^p_\alpha(t) \Delta \hat{I}^q_\beta(0) \rangle$ is the current-current time-correlator, and $\Delta \hat{I}^p_\alpha(t)=\hat{I}^p_\alpha(t)-\langle \hat I^p_\alpha\rangle$, with $\hat{I}^p_\alpha$, the electrical ($p=0$) or heat ($p=1$) current operator from the reservoir $\alpha$ to the central region through the barrier~$\alpha$. The finite-frequency non-symmetrized noise $\mathcal{S}^{pq}_{\alpha\beta}(\omega)$ quantifies the correlation between the currents $\hat{I}^{p}_{\alpha}$ and $\hat{I}^{q}_{\beta}$ at finite frequency $\omega$. Such a general definition embeds three types of noise: (i) 
the charge noise, $\mathcal{S}^{00}_{\alpha\beta}(\omega)$, corresponding to the correlator between the electrical current and itself; (ii) the mixed noises, $\mathcal{S}^{01}_{\alpha\beta}(\omega)$ and  $\mathcal{S}^{10}_{\alpha\beta}(\omega)$, corresponding to the correlators between the electrical current and the heat current, and (iii) the heat noise, $\mathcal{S}^{11}_{\alpha\beta}(\omega)$, corresponding to the correlator between the heat current and itself.

We first compute the time-correlator $\mathcal{S}^{pq}_{\alpha\beta}(t,t')$ using the Keldysh out-of-equilibrium formalism \cite{Keldysh1964}, and next calculate its Fourier transform in order to get $\mathcal{S}^{pq}_{\alpha\beta}(\omega)$. To achieve this task, we insert the current operator, given by Eq.~(\ref{c1}), in the definition of the noise, given by Eq.~(\ref{def_S}), and perform the calculation of  the average of the product of four creation/annihilation operators, making the following assumptions: non-interacting electrons, wide-band approximation, and symmetrical coupling strength between the reservoirs and the QD (i.e., symmetrical left and right barriers). The details of the calculation are given in the Appendix~\ref{appendixA}. The final expression of finite-frequency non-symmetrized noise we obtain is
\begin{eqnarray}\label{exp_noise}
&&\mathcal{S}^{pq}_{\alpha\beta}(\omega)=\frac{e^{2-p-q}}{h}\int_{-\infty}^{\infty} d\varepsilon \nonumber\\
&&
\times\Big[(\varepsilon-\mu_\alpha)^{p}(\varepsilon-\mu_\beta)^{q}\mathcal{A}_{\alpha\beta}(\varepsilon,\omega)\nonumber\\
&&+(\varepsilon-\mu_\alpha)^{p}(\varepsilon-\hbar\omega-\mu_\beta)^{q}\mathcal{B}_{\alpha\beta}(\varepsilon,\omega)\nonumber\\
&&+(\varepsilon-\hbar\omega-\mu_\alpha)^{p}(\varepsilon-\mu_\beta)^{q}\mathcal{B}^*_{\beta\alpha}(\varepsilon,\omega)\nonumber\\
&&
+(\varepsilon-\hbar\omega-\mu_\alpha)^{p}(\varepsilon-\hbar\omega-\mu_\beta)^{q}\mathcal{C}_{\alpha\beta}(\varepsilon,\omega)\Big]~,
\end{eqnarray}
with
\begin{eqnarray}\label{A}
&&\mathcal{A}_{\alpha\beta}(\varepsilon,\omega)
=\mathcal{T}(\varepsilon-\hbar\omega)f_M^h(\varepsilon-\hbar\omega)\Big[\mathcal{T}(\varepsilon)f_M^e(\varepsilon)\nonumber\\
&&+[\delta_{\alpha\beta}-t(\varepsilon)]f_\alpha^e(\varepsilon)+[\delta_{\alpha\beta}-t^*(\varepsilon)]f_\beta^e(\varepsilon)\Big]~,
\end{eqnarray}
\begin{eqnarray}\label{B}
 &&\mathcal{B}_{\alpha\beta}(\varepsilon,\omega)
=t(\varepsilon)t(\varepsilon-\hbar\omega)\left[f_\alpha^e(\varepsilon)-t^*(\varepsilon)f_M^e(\varepsilon)\right]\nonumber\\
&&\times
\left[f_\beta^h(\varepsilon-\hbar\omega)-t^*(\varepsilon-\hbar\omega)f_M^h(\varepsilon-\hbar\omega)\right]~,
\end{eqnarray}
and
\begin{eqnarray}\label{C}
\mathcal{C}_{\alpha\beta}(\varepsilon,\omega)
&=&\mathcal{T}(\varepsilon)f_M^e(\varepsilon)\Big[\mathcal{T}(\varepsilon-\hbar\omega)f_M^h(\varepsilon-\hbar\omega)\nonumber\\
&&+[\delta_{\alpha\beta}-t^*(\varepsilon-\hbar\omega)]f_\alpha^h(\varepsilon-\hbar\omega)\nonumber\\
&&+[\delta_{\alpha\beta}-t(\varepsilon-\hbar\omega)]f_\beta^h(\varepsilon-\hbar\omega)\Big]~,
\end{eqnarray}
where $f_\alpha^e(\varepsilon)=[1+\exp((\varepsilon-\mu_\alpha)/k_BT_\alpha)]^{-1}$ is the Fermi-Dirac distribution function for electrons, $f_\alpha^h(\varepsilon)=1- f_\alpha^e(\varepsilon)$ is the distribution function for holes,
$f_M^{e,h}(\varepsilon)=[f_L^{e,h}(\varepsilon)+f_R^{e,h}(\varepsilon)]/2$ is the average left and right distribution, $t(\varepsilon)$ is the transmission amplitude, and $\mathcal{T}(\varepsilon)=|t(\varepsilon)|^2$ is the transmission coefficient. The transmission amplitude is related to the retarded Green function of the QD, $G^r(\varepsilon)$, through the relation $t(\varepsilon)=i\Gamma G^r(\varepsilon)$, where $\Gamma$ is the coupling strength between the QD and the reservoirs \cite{Zamoum2016}.

Equation~(\ref{exp_noise}) gives the electrical noise when $p=q=0$, it gives the mixed noise when either $p=0$ and $q=1$, or vice versa, and it gives the heat noise when $p=q=1$. $\mathcal{S}^{pq}_{\alpha\beta}(\omega)$ is a real quantity when $p=q$ and $\alpha=\beta$ (auto-correlator), but can be complex otherwise (cross-correlator).  We have checked that the electrical noise $\mathcal{S}^{00}_{\alpha\beta}(\omega)$ extracted from Eq.~(\ref{exp_noise}) coincides with the results of the literature \cite{Hammer2011,Zamoum2016}, and that the heat noise $\mathcal{S}^{11}_{\alpha\beta}(\omega)$ extracted from Eq.~(\ref{exp_noise}) coincides with the existing results of the  literature in the limit of energy-independent transmission amplitude \cite{Sergi2011}. The expressions for the mixed noises $\mathcal{S}^{01}_{\alpha\beta}(\omega)$ and $\mathcal{S}^{10}_{\alpha\beta}(\omega)$ are novels. This is the central result of this paper. It is valid at any frequency $\omega$, coupling strength $\Gamma$ between the 
QD and the reservoirs, 
QD energy level $\varepsilon_d$, and for any temperature and voltage gradients between the left and right reservoirs.

In the following, we choose first to restrict our study to the case where temperatures for the left and right reservoirs are equal, $T_L=T_R=T$, and for $\varepsilon_d=0$ (electron-hole symmetry point), and we discuss the mixed noise spectrum in three situations: (i) at equilibrium, (ii) for energy-independent transmission amplitude, and (iii) for an Anderson-type transmission amplitude. In the latter case, we also discuss the spectral coherence in the presence of a temperature gradient between the two reservoirs.


\section{Discussion}

%
%

\subsection{At equilibrium}

At equilibrium, i.e., $\mu_L=\mu_R=\varepsilon_F$, where $\varepsilon_F$ is the Fermi energy for electrons in the reservoirs, and for equal left and right reservoir temperatures, i.e., $T_L=T_R=T$, we have from Eqs.~(\ref{exp_noise})-(\ref{C})
\begin{eqnarray}\label{S_eq}
&&\mathcal{S}^{pq}_{\alpha\beta}(\omega)=\frac{e^{2-p-q}}{h}\int_{-\infty}^{\infty} d\varepsilon f_M^e(\varepsilon)f_M^h(\varepsilon-\hbar\omega)\nonumber\\
&&
\times\Big[\varepsilon^{p+q}\mathcal{\widetilde A}_{\alpha\beta}(\varepsilon,\omega)+\varepsilon^{p}(\varepsilon-\hbar\omega)^{q}\mathcal{\widetilde B}_{\alpha\beta}(\varepsilon,\omega)\nonumber\\
&&+(\varepsilon-\hbar\omega)^{p}\varepsilon^{q}\mathcal{\widetilde B}^*_{\beta\alpha}(\varepsilon,\omega)
+(\varepsilon-\hbar\omega)^{p+q}\mathcal{\widetilde C}_{\alpha\beta}(\varepsilon,\omega)\Big]~,\nonumber\\
\end{eqnarray}
with
\begin{eqnarray}
\mathcal{\widetilde A}_{\alpha\beta}(\varepsilon,\omega)
&=&\mathcal{T}(\varepsilon-\hbar\omega)
\left[2\delta_{\alpha\beta}-\mathcal{T}(\varepsilon)\right]~,\\
\mathcal{\widetilde B}_{\alpha\beta}(\varepsilon,\omega)
&=&t(\varepsilon)t(\varepsilon-\hbar\omega)\nonumber\\
&&\times\left[1-t^*(\varepsilon)\right]\left[1-t^*(\varepsilon-\hbar\omega)\right]~,\\
\mathcal{\widetilde C}_{\alpha\beta}(\varepsilon,\omega)
&=&\mathcal{T}(\varepsilon)
\left[2\delta_{\alpha\beta}-\mathcal{T}(\varepsilon-\hbar\omega)\right]~,
\end{eqnarray}
since for isothermal reservoirs at equilibrium, we have $f^{e,h}_L(\varepsilon)=f^{e,h}_R(\varepsilon)=f^{e,h}_M(\varepsilon)$. From Eq.~(\ref{S_eq}), it can be shown using $\mathcal{S}^{pq}_{\alpha\beta}(-\omega)=e^{\hbar\omega/k_BT}\mathcal{S}^{qp}_{\alpha\beta}(\omega)$ that the noise spectrum obeys the relation
\begin{eqnarray}
\mathcal{S}^{pq}_{\alpha\beta}(\omega)=N(\hbar\omega)[\mathcal{S}^{qp}_{\alpha\beta}(-\omega)-\mathcal{S}^{pq}_{\alpha\beta}(\omega)]~,
\end{eqnarray}
where $N(\hbar\omega)=[\exp(\hbar\omega/k_BT)-1]^{-1}$ is the Bose-Einstein distribution function. Removing the reservoirs' index and using the definitions \cite{Andrade2015} of the electrical ac-conductance, $G(\omega)=[\mathcal{S}^{00}(-\omega)-\mathcal{S}^{00}(\omega)]/2\hbar\omega$, the thermal ac-conductance $K(\omega)=[\mathcal{S}^{11}(-\omega)-\mathcal{S}^{11}(\omega)]/2\hbar\omega T$, and the thermoelectric ac-conductance, $X(\omega)=[\mathcal{S}^{10}(-\omega)-\mathcal{S}^{01}(\omega)]/2\hbar\omega T$ which is the product of the ac-thermopower (i.e., Seebeck coefficient) by the electrical ac-conductance, we establish that the noises are related at equilibrium to the ac-conductances through the following fluctuation-dissipation relations
\begin{eqnarray}\label{G}
\mathcal{S}^{00}(\omega)&=&2\hbar\omega N(\hbar\omega)G(\omega)~,\\\label{X}
\mathcal{S}^{01}(\omega)&=&2\hbar\omega T N(\hbar\omega)X(\omega)~,\\\label{K}
\mathcal{S}^{11}(\omega)&=&2\hbar\omega T N(\hbar\omega)K(\omega)~.
\end{eqnarray}
Through these relations, we can state that in a similar way that the finite-frequency electrical noise contains information on the dynamics of the charge transfer, the finite-frequency heat noise contains information on the dynamic of the heat transfer (since $G(\omega)$ and $K(\omega)$ are the response to an excitation modulated in time). Moreover, Eq.~(\ref{X}) confirms the key role played by the mixed noise $\mathcal{S}^{01}(\omega)$ to quantify the thermoelectric conversion. Note that in the limit of zero-frequency, Eqs.~(\ref{G})-(\ref{K}) reduce to the relations given in Ref.~\onlinecite{Crepieux2015}, since we have in that limit $N(\hbar\omega)\rightarrow k_BT/\hbar\omega$.

%
%

\subsection{Energy-independent transmission}

\begin{figure}[h!]
\begin{center}
\includegraphics[width=8cm]{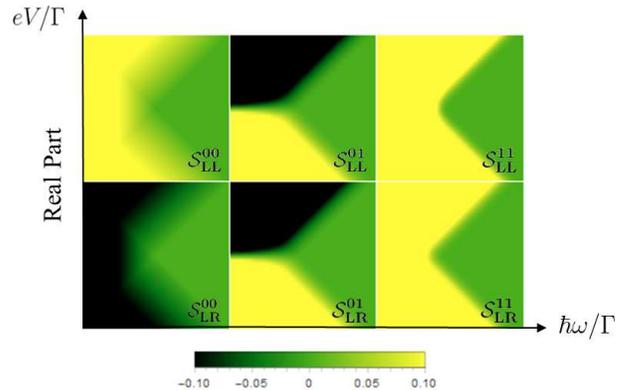}
\caption{Noises spectra as a function of frequency $\hbar\omega/\Gamma\in[-10,10]$ and voltage $eV/\Gamma\in[-10,10]$ for $\mathcal{T}=0.01$ and $t=\mathcal{T}+i[\mathcal{T}(1-\mathcal{T})]^{1/2}$, at low temperature $k_BT/\Gamma=0.01$. $\mathcal{S}^{pq}_{\alpha\beta}(\omega)$ is plotted in units of $e^{2-p-q}\Gamma^{1+p+q}/h$. The right reservoir is grounded ($\mu_R=0$).}
\label{figure2}
\end{center} 
\end{figure}

For an energy-independent transmission amplitude, $t$, the real parts of the electrical, mixed, and heat noise spectra are given by Fig.~\ref{figure2} in the low-temperature limit. We do not plot their imaginary parts whose magnitudes are smaller with a factor 100 comparing to the ones of the real parts. Let us now discuss the features appearing on Fig.~\ref{figure2}. First, we notice that similarly to the electrical noise, which cancels when the frequency is larger than the voltage, $\hbar\omega>eV$, the mixed and heat noises cancel as well. The reason is the following: knowing that the noise is called emission noise at positive frequency and absorption noise at negative frequency \cite{Basset2010}, we understand that the system can not emit an energy larger than the energy provided to it, here the voltage since temperature is taken small. Second, we observe that the electrical noise varies linearly or by plateaus with both voltage and frequency, due to the fact that when transmission is energy independent, the system works in the linear regime. 
Third, the mixed noise can change its sign whereas the electrical and heat noises keep a single sign. Fourth, the electrical and mixed correlators between distinct reservoirs, $\mathcal{S}^{pq}_{LR}$, are equal in absolute values to the correlators in the same reservoir, $\mathcal{S}^{pq}_{LL}$, and nearly equal for the heat correlator \cite{note0}.

In the limit of weak or perfect transmission, i.e., $\mathcal{T}\ll 1$ or $\mathcal{T}=1$ respectively, the integration over energy in Eq.~(\ref{exp_noise}) can be performed analytically (see Appendix~\ref{appendixB} for the details of the calculation). The expressions of the noises, which are all real in these limits, are given in Table~\ref{table1}. These expressions constitute a generalization of the fluctuation-dissipation theorem to an out-of-equilibrium situation since the Bose-Einstein distribution function is estimated at frequency shifted by $\pm eV/\hbar$. Concerning the electrical noise, its expression at $\mathcal{T}\ll 1$ is in full agreement with the result of perturbative calculations \cite{Roussel2016}. Concerning the mixed and heat noises expressions, there is no previous work to compare in the literature. Note that at zero-voltage, the mixed noise cancels in both limits ($\mathcal{T}\ll 1$ and $\mathcal{T}=1$) but not in the intermediate regime: the mixed noise is then given by Eq. (\ref{X})
 which is a priori non-zero. It is also worth to notice that $\mathcal{S}^{11}_{\alpha\beta}(\omega)$ contains a contribution which is proportional to $\mathcal{S}^{00}_{\alpha\beta}(\omega)$ with a proportionality factor equal to $\mathcal{L}T^2$, where $\mathcal{L}=\pi^2k_B^2/3e^2$ is the Lorenz number. Since in certain limits, the heat noise is related to the thermal conductance and the electrical noise to the electrical conductance, as through Eqs.~(\ref{K}) and (\ref{G}) at equilibrium for example, it is not surprising to find a relation which involves the Lorenz number between the heat noise and the electrical noise thanks to the Wiedemann-Franz law, or between the thermal conductance and the electrical noise as obtained in Ref.~\onlinecite{Gnezdilov2016}. Table~\ref{table1} gives also the sum over reservoirs of the electrical, mixed, and heat noises, $\sum_{\alpha\beta}\mathcal{S}^{pq}_{\alpha\beta}(\omega)$. Contrary to the total electrical and mixed noises, which are equal to zero in the limits we 
consider (no charging effect on the QD), the total heat noise takes a finite value which indeed corresponds to the heat power fluctuations. At zero-frequency, the power fluctuations are conserved, i.e., the heat power fluctuations are equal to the electrical power fluctuations \cite{Crepieux2015}. It is also true at finite-frequency provided that $\mathcal{T}=1$, since in that limit we have from Table~\ref{table1}, $\sum_{\alpha\beta}\mathcal{S}^{11}_{\alpha\beta}(\omega)=V^2\mathcal{S}^{00}_{\alpha\alpha}(\omega)$. At zero-temperature, we get for $\mathcal{T}=1$: $\sum_{\alpha\beta}\mathcal{S}^{11}_{\alpha\beta}(\omega)=2|\hbar\omega|(eV)^2\Theta(-\omega)/h$, and for $\mathcal{T}\ll 1$, $\sum_{\alpha\beta}\mathcal{S}^{11}_{\alpha\beta}(\omega)=4\mathcal{T}|\hbar\omega|^3\Theta(-\omega)/h$ at zero-voltage, and $\sum_{\alpha\beta}\mathcal{S}^{11}_{\alpha\beta}(\omega)=\mathcal{T}|eV|^3/h$ at zero-frequency in good agreement with Ref.~\onlinecite{Sergi2011}.

\begin{widetext}

\begin{table}[h!]
\begin{center}
\begin{tabular}{|l|c|c|c|}
\hline
Type of noise&Notation& $\mathcal{T}\ll 1$&$\mathcal{T}=1$\\ \hline\hline
Electrical noise&$\mathcal{S}^{00}_{\alpha\beta}(\omega)$& $\frac{e^2\mathcal{T}}{h}(2\delta_{\alpha\beta}-1)\sum_{\pm}(\hbar\omega\pm eV)N(\hbar\omega\pm eV)$&$\frac{e^2}{h}(2\delta_{\alpha\beta}-1)2\hbar\omega N(\hbar\omega)$\\ 
\hline
Total electrical noise& $\sum_{\alpha\beta}\mathcal{S}^{00}_{\alpha\beta}(\omega)$& $0$&$0$\\ 
 \hline
Mixed noise&$\mathcal{S}^{01}_{\alpha\beta}(\omega)$& $\frac{e\mathcal{T}}{h}(2\delta_{\alpha L}-1)\sum_{\pm}\mp\frac{(\hbar\omega\pm eV)^2}{2}N(\hbar\omega\pm eV)$&$\frac{e}{h}(1-2\delta_{\alpha L})eV\hbar\omega N(\hbar\omega)$\\ 
\hline
Total mixed noise&  $\sum_{\alpha\beta}\mathcal{S}^{01}_{\alpha\beta}(\omega)$&$0$&$0$\\ 
\hline
Heat noise&$\mathcal{S}^{11}_{\alpha\alpha}(\omega)$& $\mathcal{L}T^2\mathcal{S}^{00}_{\alpha\alpha}(\omega)$ & $\left(\mathcal{L}T^2+\frac{V^2}{2}+\frac{\hbar^2\omega^2}{12e^2}\right)\mathcal{S}^{00}_{\alpha\alpha}(\omega)$\\ 
\it{(auto-correlator)}&&$+\frac{\mathcal{T}}{h}\left[\hbar^3\omega^3N(\hbar\omega)+\sum_{\pm}\frac{(\hbar\omega\pm eV)^3}{3}N(\hbar\omega\pm eV)\right]$&$+\frac{\hbar^2\omega^2}{4h}\sum_{\pm}(\hbar\omega\pm eV)N(\hbar\omega\pm eV)$\\
 \hline
Heat noise&$\mathcal{S}^{11}_{\alpha\bar\alpha}(\omega)$& $\mathcal{L}T^2\mathcal{S}^{00}_{\alpha\bar\alpha}(\omega)$&$\mathcal{L}T^2\mathcal{S}^{00}_{\alpha\bar\alpha}(\omega)-\frac{\hbar^2\omega^2}{12e^2}\mathcal{S}^{00}_{\alpha\alpha}(\omega)$\\
\it{(cross-correlator)}&&$+\frac{\mathcal{T}}{h}\sum_{\pm}\frac{(\hbar\omega\pm eV)^3}{6}N(\hbar\omega\pm eV)$&$-\frac{\hbar^2\omega^2}{4h}\sum_{\pm}(\hbar\omega\pm eV)N(\hbar\omega\pm eV)$\\
 \hline
Total heat noise& $\sum_{\alpha\beta}\mathcal{S}^{11}_{\alpha\beta}(\omega)$&$\frac{\mathcal{T}}{h}\left[2\hbar^3\omega^3N(\hbar\omega)+\sum_{\pm}(\hbar\omega\pm eV)^3N(\hbar\omega\pm eV)\right]$&$V^2\mathcal{S}^{00}_{\alpha\alpha}(\omega)$\\ 
\hline
\end{tabular}
\caption{Expressions of the electrical, mixed, and heat noises in the energy-independent weak/perfect transmission limits \cite{note1}. We have $\bar\alpha=R$ when $\alpha=L$, and vice versa. The total electrical, mixed and heat noises summed over both reservoirs are also given. The details of the calculations are performed in Appendix B.}
\label{table1}
\end{center}
\end{table}

\end{widetext}

%
%

\subsection{Anderson-type energy transmission}

\begin{figure}[h!]
\begin{center}
\includegraphics[width=8cm]{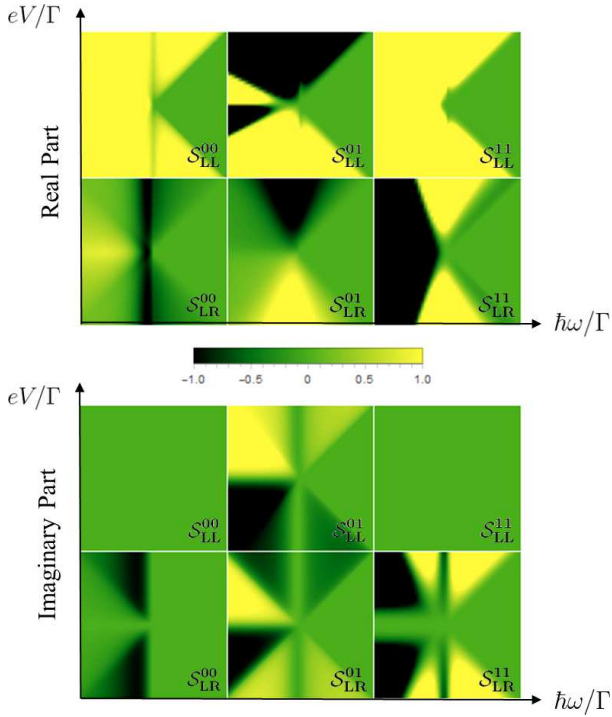}
\caption{Noises spectra as a function of frequency $\hbar\omega/\Gamma\in[-10,10]$ and voltage $eV/\Gamma\in[-10,10]$ for $\mathcal{T}(\varepsilon)=\Gamma^2/(\varepsilon^2+\Gamma^2)$, at low temperature $k_BT/\Gamma=0.01$. $\mathcal{S}^{pq}_{\alpha\beta}(\omega)$ is plotted in units of $e^{2-p-q}\Gamma^{1+p+q}/h$. The right reservoir is grounded ($\mu_R=0$).}
\label{figure3}
\end{center} 
\end{figure}

For an Anderson-type transmission amplitude of the form $t(\varepsilon)=i\Gamma/[(\varepsilon-\varepsilon_d)+i\Gamma]$, both the real and imaginary parts of the electrical, mixed, and heat noise spectra are given by Fig.~\ref{figure3} in the low temperature limit. Note that the imaginary parts of $\mathcal{S}^{00}_{LL}(\omega)$ and $\mathcal{S}^{11}_{LL}(\omega)$ are both zero since the auto-correlators are real quantities, and that the real and imaginary parts of the cross-correlators are of the same order of magnitude, contrary to the energy independent transmission case. The main observation is the dramatically distinct spectra that we have for the auto-correlators, $\mathcal{S}^{00}_{LL}(\omega)$ and $\mathcal{S}^{11}_{LL}(\omega)$, in comparison to the cross-correlators, $\mathcal{S}^{pq}_{\alpha\beta}(\omega)$ with $p\ne q$ or/and $\alpha\ne\beta$. Whereas the auto-correlator spectra are quite similar to the ones obtained in the case of an energy-independent transmission amplitude (compare to Fig.~\ref{figure2}), excepted an additional structure in the region of small positive frequency, the cross-correlator spectra exhibit the following features: (i) their sign can change, (ii) it appears a new region with specific behavior close to small frequency, but (iii) we still have a cancellation of the noises for $\hbar\omega>eV$, again due to the fact that the system can not emit energy larger than the one provided to it. We remark that in any 
situations, those depicted in Figs.~\ref{figure2} and \ref{figure3} and those summarized in Table~I, the mixed noise cancels at zero-voltage, meaning that the 
cancellation of the ratio $-V/\Delta T$, which is equal to the Seebeck coefficient $S_T$ for open circuit, causes the cancellation of the mixed noise. This is one evidence that thermopower and mixed noise are closely connected.

%
%

\begin{figure}[h!]
\begin{center}
\includegraphics[width=4cm]{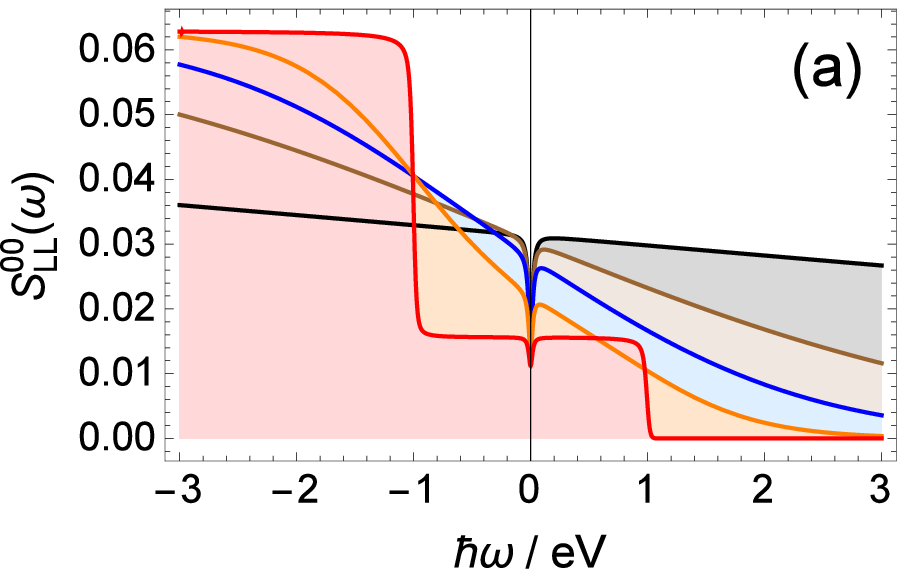}
\includegraphics[width=4cm]{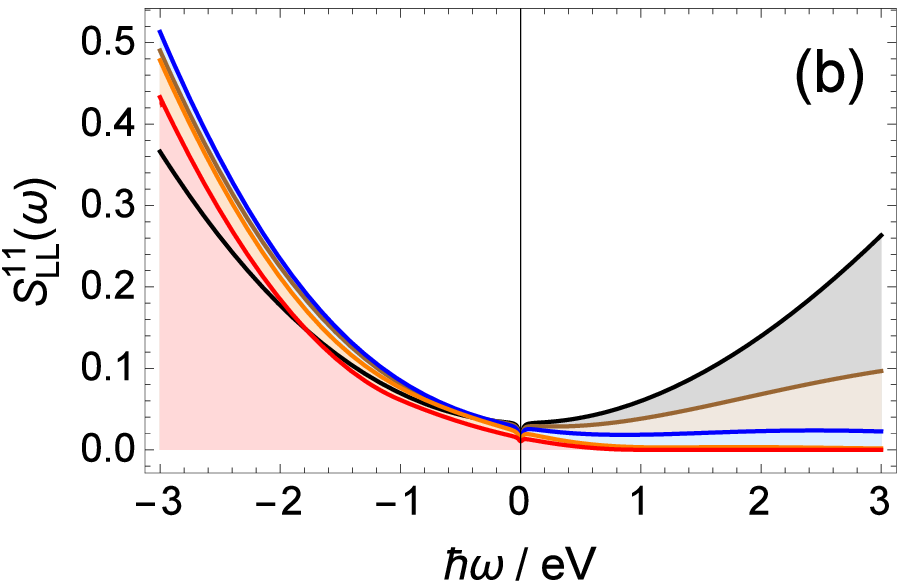}
\includegraphics[width=4cm]{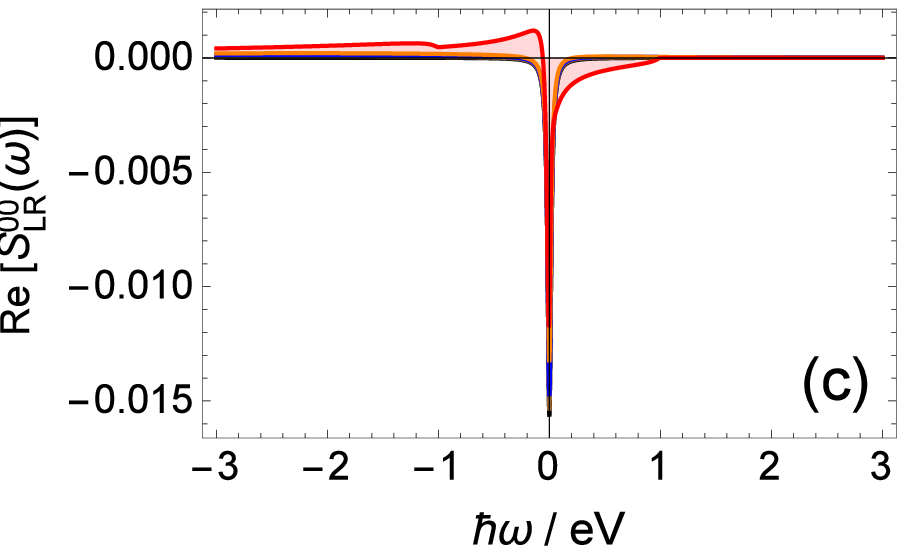}
\includegraphics[width=4cm]{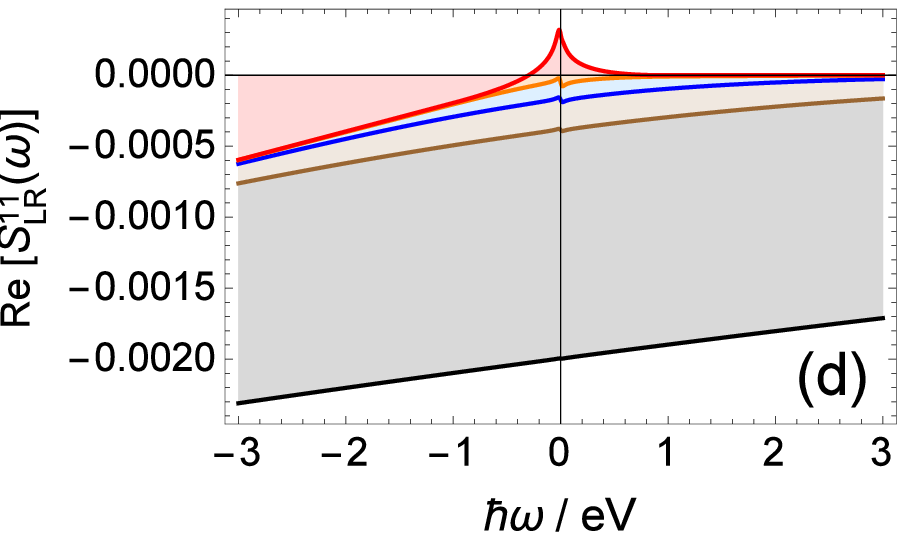}
\includegraphics[width=4cm]{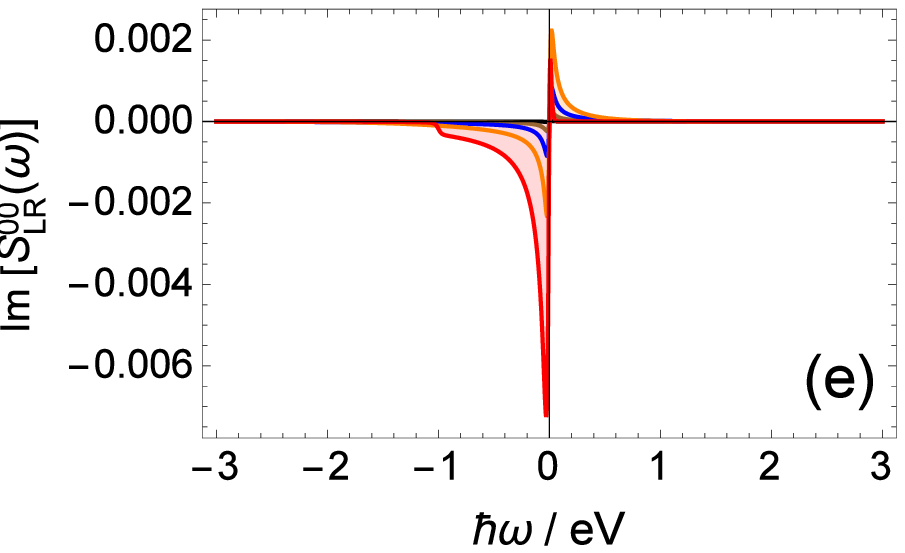}
\includegraphics[width=4cm]{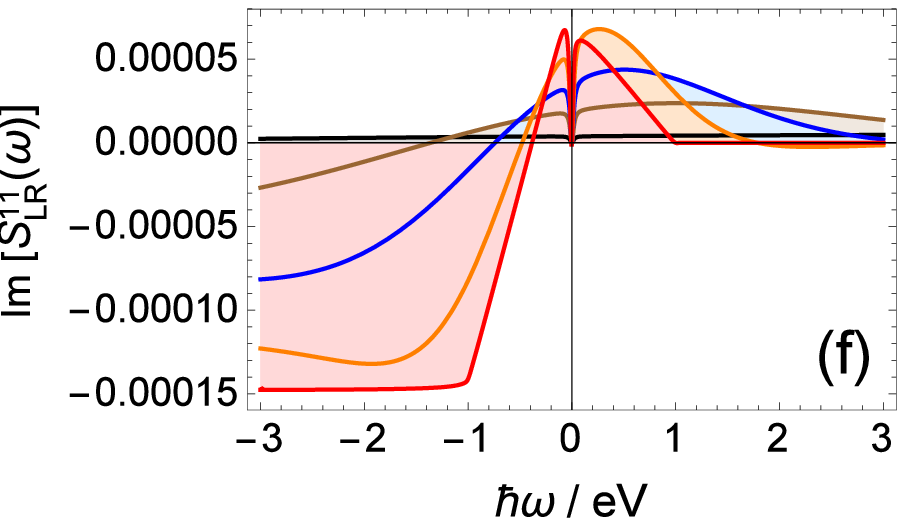}
\caption{Electrical noises $\mathcal{S}^{00}_{LL}(\omega)$ and $\mathcal{S}^{00}_{LR}(\omega)$ (left column) and heat noises $\mathcal{S}^{11}_{LL}(\omega)$ and $\mathcal{S}^{11}_{LR}(\omega)$ (right column), as a function of frequency $\hbar\omega/eV$, for $\mathcal{T}(\varepsilon)=\Gamma^2/(\varepsilon^2+\Gamma^2)$, with $\Gamma/eV=0.01$, and for varying values of the temperature $k_BT/eV$: 0.01 (red line), 0.5 (orange line), 1 (blue line), 2 (brown line), and 10 (black line). $\mathcal{S}^{pq}_{\alpha\beta}(\omega)$ is plotted in units of $e^{2-p-q}(eV)^{1+p+q}/h$. The right reservoir is grounded ($\mu_R=0$).}
\label{figure4}
\end{center} 
\end{figure}

\begin{figure}[h!]
\begin{center}
\includegraphics[width=4cm]{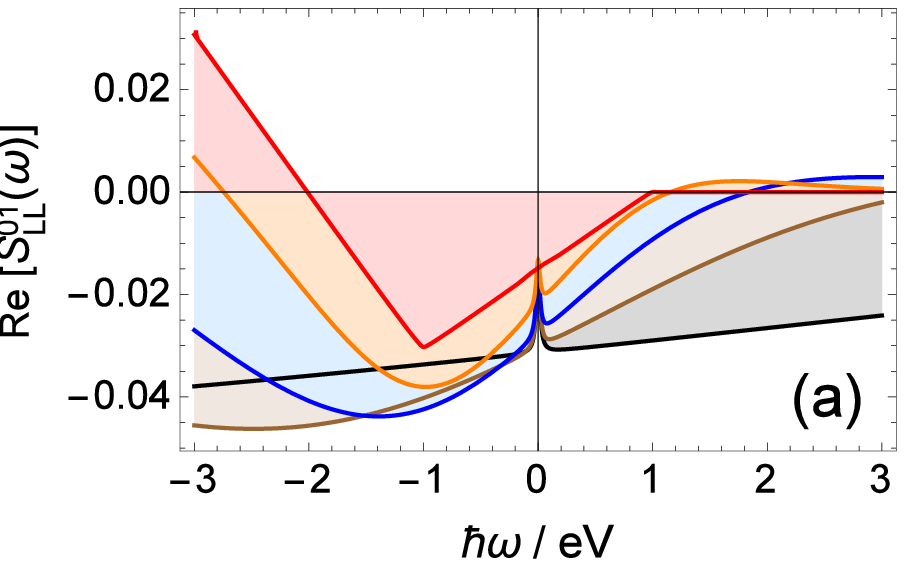}
\includegraphics[width=4cm]{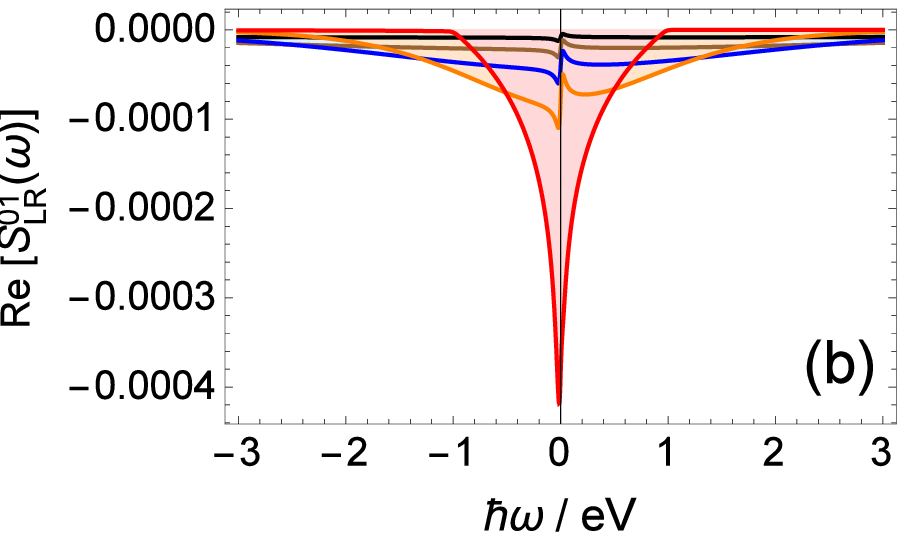}
\includegraphics[width=4cm]{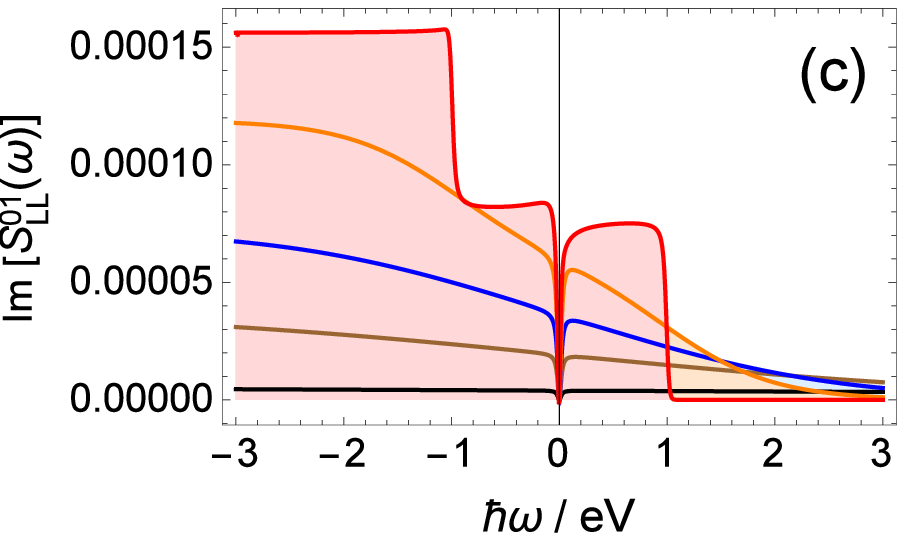}
\includegraphics[width=4cm]{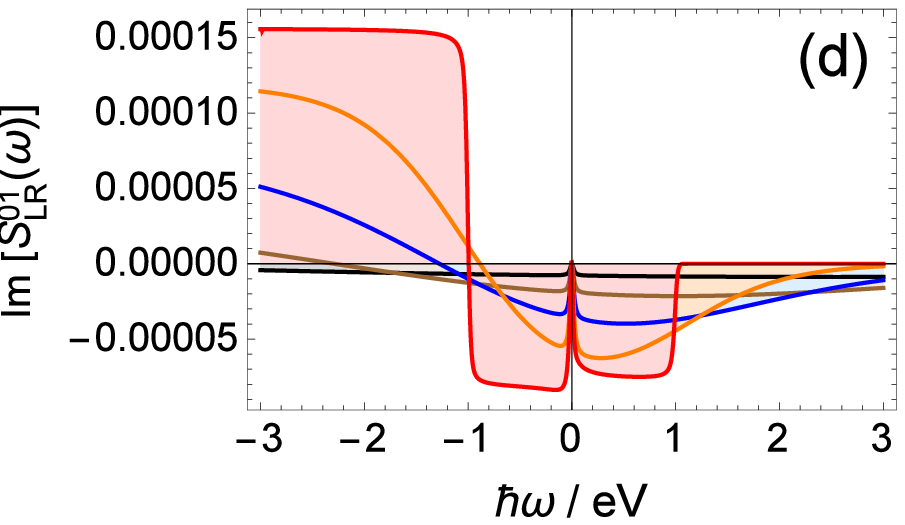}
\caption{Mixed noises $\mathcal{S}^{01}_{LL}(\omega)$ (left column) and $\mathcal{S}^{01}_{LR}(\omega)$ (right column), as a function of frequency $\hbar\omega/eV$. Same parameters as in Fig.~\ref{figure4}.}
\label{figure5}
\end{center} 
\end{figure}

To have a deeper insight in the electrical, mixed, and heat noises, we plot their real and imaginary parts as a function of frequency at weak coupling strength $\Gamma$, for increasing temperatures in Figs.~\ref{figure4} and \ref{figure5}. All the types of noise exhibit an asymmetric spectrum at low temperature (red curves) and a nearly symmetrical spectrum at large temperature with a vanishing imaginary part (black curves), due to the fact that when the temperature increases we are leaving the quantum regime. Thus, at large temperature, it is no longer necessary to make the distinction between non-symmetrized and symmetrized noises since the currents are no longer operators but just scalars (classical regime). The electrical and heat auto-correlators (see Figs.~\ref{figure4}(a) and \ref{figure4}(b)) are real and positive quantities. The electrical auto-correlator, $\mathcal{S}^{00}_{LL}(\omega)$, is strongly frequency dependent at low temperature with a down staircase-like behavior starting from the value $2\pi\Gamma e^2/h$ and going to the value $0$  (see red curve in Fig.~\ref{figure4}(a)), but resembles to a white noise at large temperature \cite{noteHT} with a constant value equals to $\pi\Gamma e^2/h$, except in a narrow low frequency region (see black curve in Fig.~\ref{figure4}(a)). At large temperature, the heat auto-correlator, $\mathcal{S}^{11}_{LL}(\omega)$, presents a power-law variation with frequency, given by $\hbar^2\omega^2\pi\Gamma/h$  (see black curve in Fig.~\ref{figure4}(b)) whereas the real part of $\mathcal{S}^{11}_{LR}(\omega)$ decreases linearly with temperature \cite{noteHT}. The electrical and heat cross-correlators, depicted in Figs.~\ref{figure4}(c)-\ref{figure4}(f), are complex quantities whose imaginary parts cancel at large temperature (black curves), making the cross-correlators real quantities in that limit. The electrical auto-correlator and the real part of the electrical cross-correlator have distinct profiles but coincide at zero-frequency in absolute value since due to charge conservation we have $\mathcal{S}^{00}_{LL}(\omega=0)=-\mathcal{S}^{00}_{LR}(\omega=0)=\pi\Gamma e^2/2h$ in that limit.

We turn now our interest to the mixed noise depicted in Fig.~\ref{figure5}. Similarly than for electrical and heat noises, increasing the temperature changes the mixed spectrum from an asymmetric profile to a symmetric profile with frequency, and cancels its imaginary part, again due to the fact that we are leaving the quantum regime. At low temperature, we also see that the imaginary parts of the mixed noises, $\mathcal{S}^{01}_{LL}(\omega)$ and $\mathcal{S}^{01}_{LR}(\omega)$, have a staircase-like profile which is a reminiscent of the electrical noise auto-correlator (compare Figs.~\ref{figure5}(c) and \ref{figure5}(d) to Fig.~\ref{figure4}(a)). Besides, the real parts of the mixed noises present quite particular profiles at low temperature: a linear profile in frequency for  $\mathcal{S}^{01}_{LL}(\omega)$ (see the red curve in Fig.~\ref{figure5}(a)) and vanishing value when $|\hbar\omega|>|eV|$ for $\mathcal{S}^{01}_{LR}(\omega)$ (see the red curve in Fig.~\ref{figure5}(b)). At large temperature, $\mathcal{S}^{01}_{LL}(\omega)$ becomes frequency independent with an asymptotic value equal to $-\pi\Gamma e^2V/h$, and $\mathcal{S}^{01}_{LR}(\omega)$ cancels \cite{noteHT}. Here again, we find that the mixed noise is related to the Seebeck coefficient $S_T$, since both quantities vary linearly with voltage.

\begin{figure}[h!]
\begin{center}
\includegraphics[width=4cm]{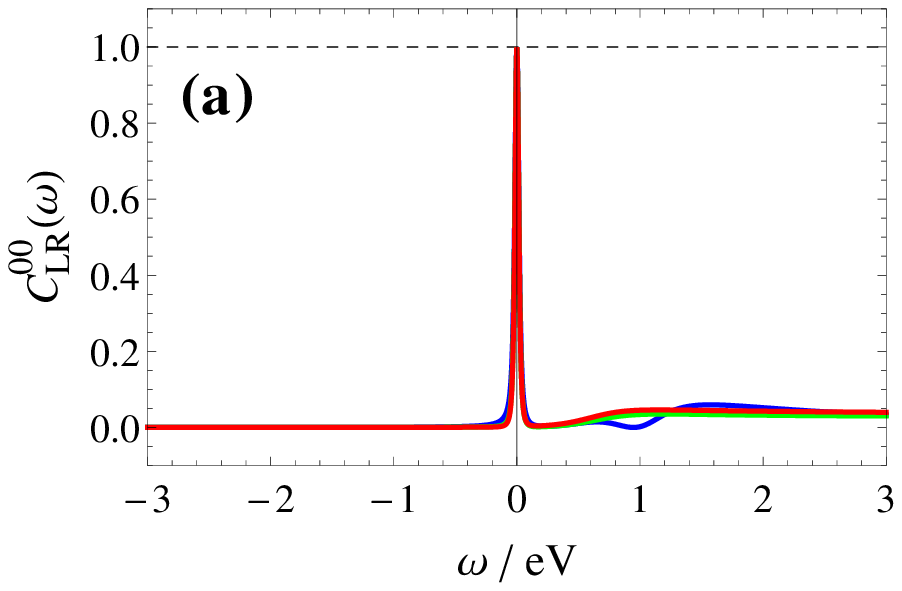}
\includegraphics[width=4cm]{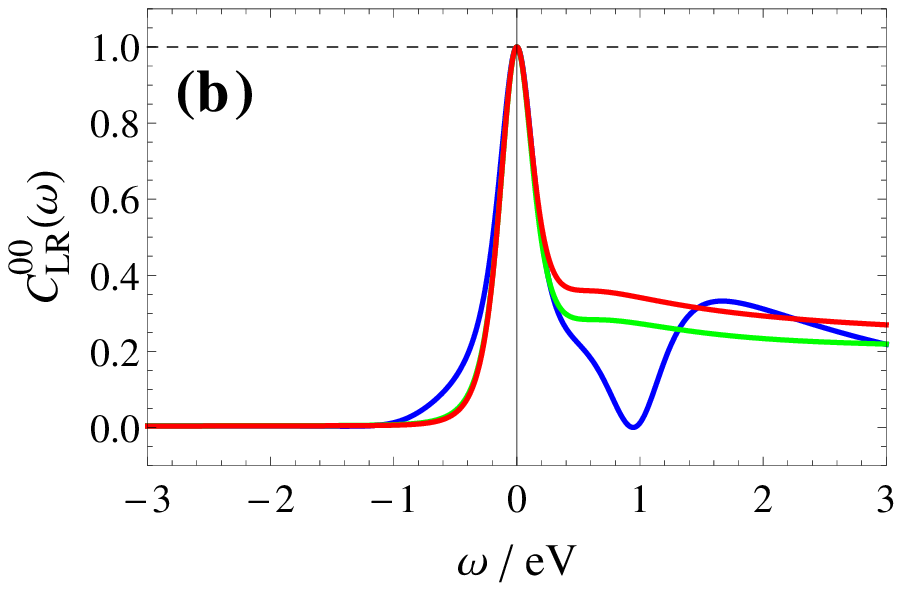}
\includegraphics[width=4cm]{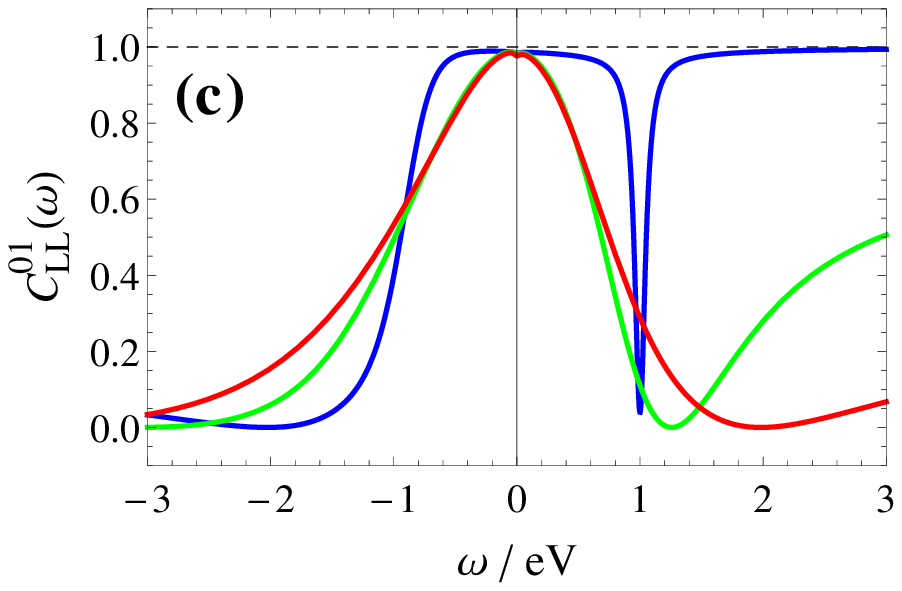}
\includegraphics[width=4cm]{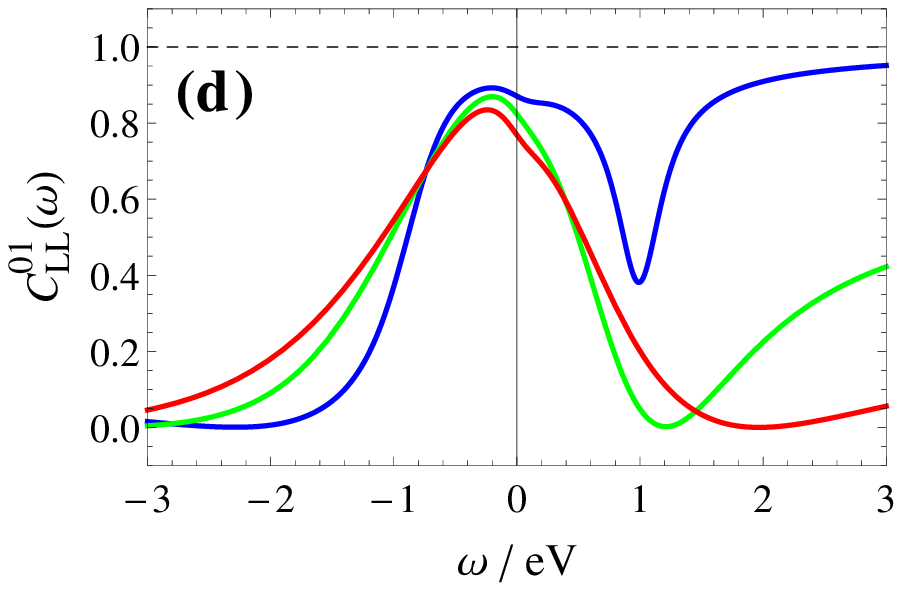}
\includegraphics[width=4cm]{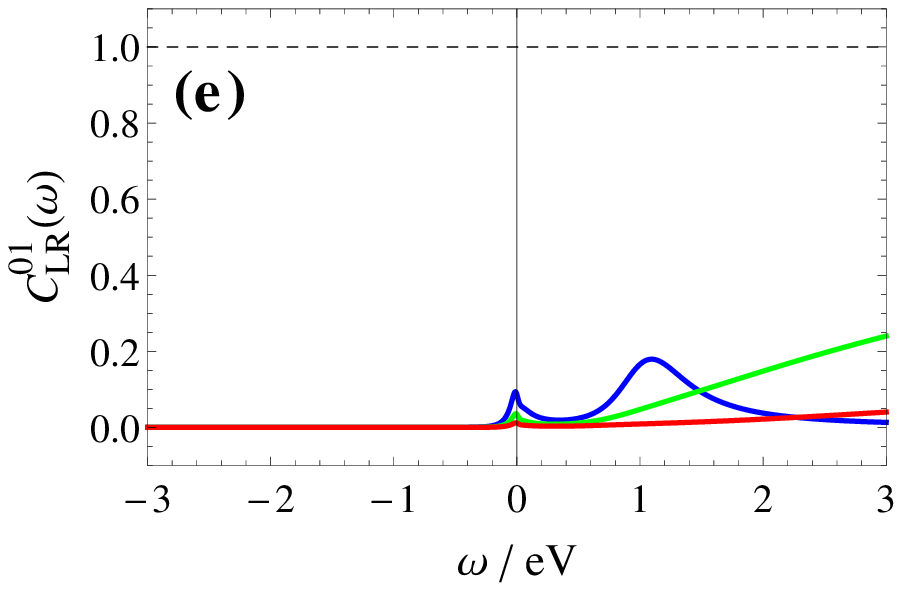}
\includegraphics[width=4cm]{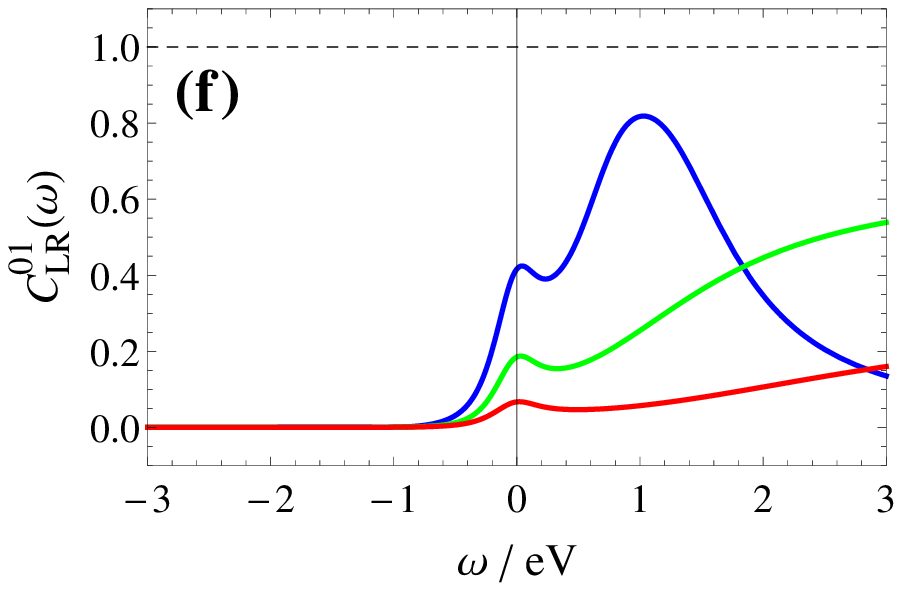}
\includegraphics[width=4cm]{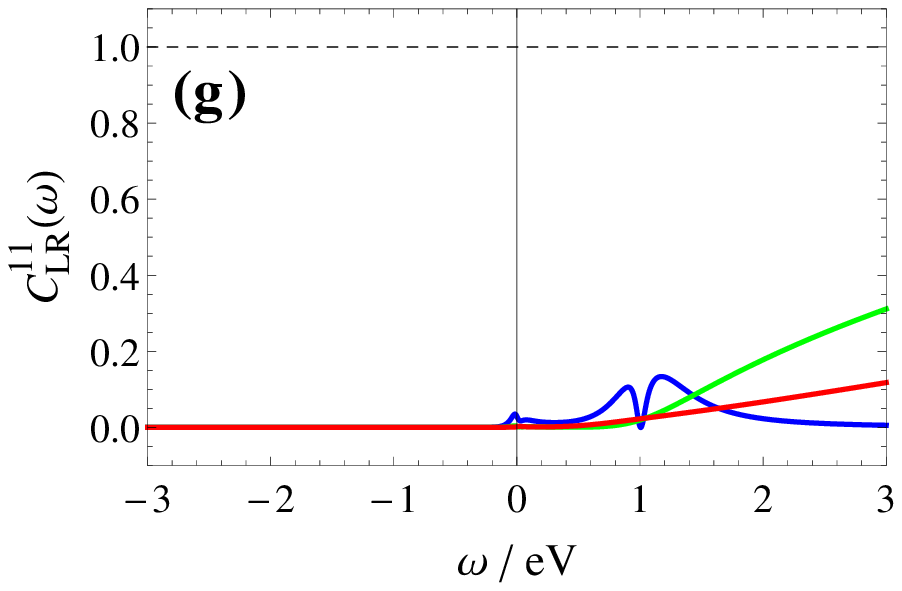}
\includegraphics[width=4cm]{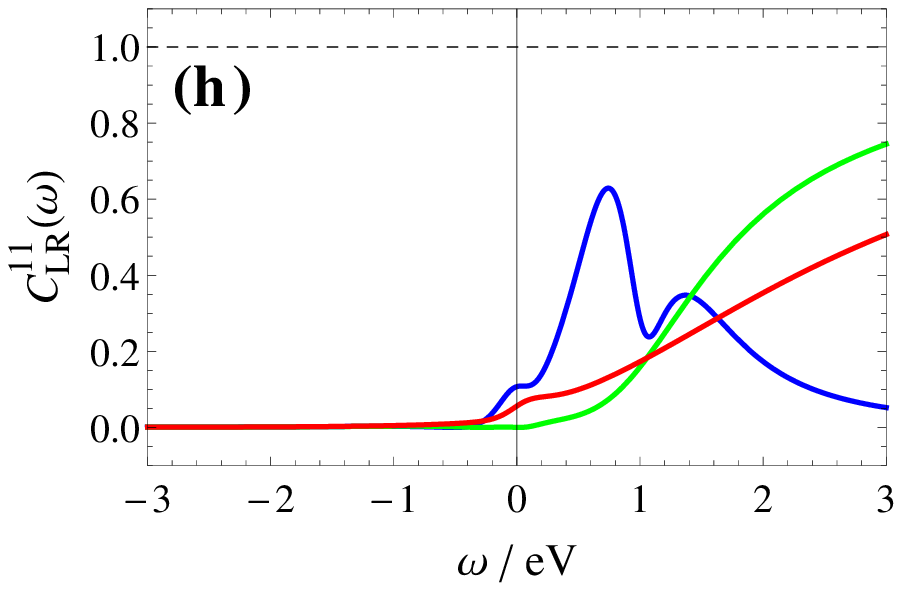}
\caption{Spectral coherence for $\mathcal{T}(\varepsilon)=\Gamma^2/(\varepsilon^2+\Gamma^2)$, for $\Gamma/eV=0.02$ (left column) and $\Gamma/eV=0.2$ (right column), at $k_BT_R/eV=0.1$ and temperature gradient equals to: $\Delta T/eV=0$ (blue curve), $\Delta T/eV=0.5$ (green curve), and $\Delta T/eV=1$ (red curve). The dashed black line shows the maximal possible value for the spectral coherence, i.e., 1.}
\label{figure6}
\end{center} 
\end{figure}

For completeness, we discuss the spectral coherence of the cross-correlators, defined as $C^{pq}_{\alpha\beta}(\omega)=|\mathcal{S}^{pq}_{\alpha\beta}(\omega)|^2/[\mathcal{S}^{pp}_{\alpha\alpha}(\omega)\mathcal{S}^{qq}_{\beta\beta}(\omega)]$, and plot their profiles on Fig.~\ref{figure6}. Thanks to Cauchy-Schwarz inequality, we have $0\le C^{pq}_{\alpha\beta}(\omega)\le 1$, where the value zero for the spectral coherence means that the currents $I^p_\alpha$ and $I^q_\beta$ are uncorrelated, whereas the value~one means that the currents $I^p_\alpha$ and $I^q_\beta$ are fully correlated. The plots on the left column of Fig.~\ref{figure6} are obtained for a weak coupling strength ($\Gamma/eV=0.02$), whereas the plots on the right column correspond to an intermediate coupling strength ($\Gamma/eV=0.2$). In the weak coupling strength limit, we remark that $C^{pq}_{\alpha\ne \beta}(\omega)$ is equal to zero at negative frequency, meaning that the absorbed signals in distinct reservoirs are uncorrelated. Moreover, we see in Fig.~\ref{figure6}(a) that the left and right electrical currents are well correlated only at zero-frequency, i.e., 
in the large-time limit, due to charge conservation which imposes $C^{00}_{LR}(\omega=0)=C^{00}_{LL}(\omega=0)=1$. When the coupling strength increases, the spectral coherence $C^{00}_{LR}(\omega)$ is broadened to non-zero frequencies (see Fig.~\ref{figure6}(b)) and can even reach $40\%$ at positive frequencies, an effect which is amplified around $\hbar\omega=eV$ when a temperature gradient is applied (see red curve in Fig.~\ref{figure6}(b)). The electrical and heat currents inside a single reservoir, here $L$, are well correlated at $\Delta T=0$  (see blue curves in Figs.~\ref{figure6}(c) and \ref{figure6}(d)) provided that $\hbar\omega >-eV$, excepted in a narrow region around $\hbar\omega=eV$ where a minimum of $C^{01}_{LL}(\omega)$ is observed. At the same frequency, $C^{01}_{LR}(\omega)$ exhibits a maximum (see blue curves in Figs.~\ref{figure6}(e) and \ref{figure6}(f)) meaning that the left electrical current and the right heat current are maximally correlated in that region of frequency. The increasing of the coupling strength reinforces this effect with values of $C^{01}_{LR}(\omega)$  up to $80\%$ (see blue curve in Fig.~\ref{figure6}(f)). This result allows us to make the prediction that the thermoelectric conversion could be optimal when the voltage applied to the QD is time-modulated with a frequency equals to the dc-voltage, i.e., $eV/\hbar$. This effect is, however, suppressed in the presence of a temperature gradient (see green and red curves in Fig.~\ref{figure6}(f)). On the contrary, $C^{11}_{LR}(\omega)$ can be increased at high frequency by the application of a temperature gradient (see Fig.~\ref{figure6}(h)).


\section{Conclusion}

We have used the Keldysh out-of-equilibrium Green function technique to calculate the finite-frequency mixed noise, and have shown that its spectrum presents a rich and specific profile, which differs from the ones of the electrical and heat noises. At equilibrium, it is related to the thermoelectric ac-conductance, meaning that the finite-frequency mixed noise gives access to the dynamics of the thermoelectric conversion. Out-of-equilibrium, by a careful study of the spectral coherence, we find that the electrical current in one reservoir is strongly correlated to the heat current in the other reservoir when the frequency is of the order of the applied voltage. The method developed here constitutes an adequate framework which can be used in future works on this quantity in more complicated quantum systems, including multi-terminals, multi-channels, and interactions in some extent.

\acknowledgments

We would like to thank M.~Guigou, M.~Lavagna, T.~Martin,
F.~Michelini, and R.~Zamoum for useful discussions. We
acknowledge financial support from the CNRS Cellule Energie
funding project ICARE, and from A*Midex.


\appendix

\section{Correlators of charge and heat currents in a QD}
\label{appendixA}

\subsection{Computation of the time-correlator $\mathcal{S}^{pq}_{\alpha\beta}(t,t')$}

We use an approach analog to the one developed by Haug and Jauho for the calculation of electrical noise \cite{Haug2007}, but instead of calculating the symmetrized noise, we calculate the non-symmetrized one since it is this latter quantity which is accessible in the experiments measuring the electrical current noise. We perform this calculation for each type of noise, the electrical, mixed and heat ones, using the general framework exposed in this Appendix.

\subsubsection{Expression of $\mathcal{S}^{pq}_{\alpha\beta}(t,t')$ in terms of the two-particle Green function of the QD, $G^{dd}_i$}

We report Eq.~(\ref{c1}) in Eq.~(\ref{def_S}) and get
\begin{eqnarray}
&&\mathcal{S}^{pq}_{\alpha\beta}(t,t')=-\frac{e^{2-p-q}}{\hbar^{2}}\sum_{\substack{k\in\alpha,k'\in\beta}}\left(\varepsilon_{k\alpha}-\mu_{\alpha}\right)^{p}\left(\varepsilon_{k'\beta}-\mu_{\beta}\right)^{q}\nonumber\\
&&\times\Big[V_{k\alpha}V_{k'\beta}\langle c^{\dag}_{k\alpha}(t)d(t)c^{\dag}_{k'\beta}(t')d(t')\rangle\nonumber\\
&&
-V_{k\alpha}V^{*}_{k'\beta}\langle c^{\dag}_{k\alpha}(t)d(t)d^{\dag}(t')c_{k'\beta}(t')\rangle\nonumber\\
&&-V^{*}_{k\alpha}V_{k'\beta}\langle d^{\dag}(t)c_{k\alpha}(t)c^{\dag}_{k'\beta}(t')d(t')\rangle\nonumber\\
&&
+V^{*}_{k\alpha}V^{*}_{k'\beta}\langle d^{\dag}(t)c_{k\alpha}(t)d^{\dag}(t')c_{k'\beta}(t')\rangle\Big]-\langle\hat{I}^{p}_{\alpha}\rangle\langle\hat{I}^{q}_{\beta}\rangle~.
\end{eqnarray}

Defining the following {\it greater} two-particle Green functions \cite{Haug2007} 
\begin{eqnarray*}
G^{cd,>}_{1}(t,t')&=&i^{2}\langle Tc^{\dag}_{k\alpha}(t)d(t)c^{\dag}_{k'\beta}(t')d(t')\rangle~,\\
G^{cd,>}_{2}(t,t')&=&i^{2}\langle Tc^{\dag}_{k\alpha}(t)d(t)d^{\dag}(t')c_{k'\beta}(t')\rangle~,\\
G^{cd,>}_{3}(t,t')&=&i^{2}\langle Td^{\dag}(t)c_{k\alpha}(t)c^{\dag}_{k'\beta}(t')d(t')\rangle~,\\
G^{cd,>}_{4}(t,t')&=&i^{2}\langle Td^{\dag}(t)c_{k\alpha}(t)d^{\dag}(t')c_{k'\beta}(t')\rangle~,
\end{eqnarray*}
and using the Keldysh formalism \cite{Keldysh1964}, the non-equilibrium contour-ordered counterparts of the correlation function can be expressed in terms of $G^{cd}_{i}(\tau,\tau')$, the contour-ordered counterparts of the two-particle Green functions, $G^{cd,>}_{i}(t,t')$, through
\begin{eqnarray}\label{g1}
&&\mathcal{S}^{pq}_{\alpha\beta}(\tau,\tau')=\frac{e^{2-p-q}}{\hbar^{2}}\sum_{\substack{k\in\alpha,k'\in\beta}}\left(\varepsilon_{k\alpha}-\mu_{\alpha}\right)^{p}\left(\varepsilon_{k'\beta}-\mu_{\beta}\right)^{q}\nonumber\\
&&\times\Big[V_{k\alpha}V_{k'\beta}G^{cd}_{1}(\tau,\tau')-V_{k\alpha}V^{*}_{k'\beta}G^{cd}_{2}(\tau,\tau')\nonumber\\
& &-V^{*}_{k\alpha}V_{k'\beta}G^{cd}_{3}(\tau,\tau')+V^{*}_{k\alpha}V^{*}_{k'\beta}G^{cd}_{4}(\tau,\tau')\Big]-\langle\hat{I}^{p}_{\alpha}\rangle\langle\hat{I}^{q}_{\beta}\rangle~.\nonumber\\
\end{eqnarray}

The next step is to express the two-particle Green functions, $G^{cd}_i$, mixing $c$ and $d$ operators in terms of the two-particle Green functions of the QD, $G^{dd}_i$, and of the bare Green function of the reservoirs, $g_{k\alpha}$. We have \cite{Haug2007}

 \begin{eqnarray*}
&&G^{cd}_{1}(\tau,\tau')=-\frac{V^{*}_{k\alpha}V^{*}_{k'\beta}}{\hbar^2}\iint d\tau_1 d\tau_2 \nonumber\\
&&\times g_{k\alpha}(\tau_{1},\tau)
 g_{k'\beta}(\tau_{2},\tau')G^{dd}_{1}(\tau,\tau',\tau_{1},\tau_{2})~,\\
&&G^{cd}_{2}(\tau,\tau')=-\delta_{kk'}\delta_{\alpha\beta}g_{k\alpha}(\tau',\tau)G(\tau,\tau')
-\frac{V^{*}_{k\alpha}V_{k'\beta}}{\hbar^2}\nonumber\\
&&\times\iint d\tau_1 d\tau_2g_{k\alpha}(\tau_{2},\tau)g_{k'\beta}(\tau',\tau_{1})G^{dd}_{2}(\tau,\tau',\tau_{1},\tau_{2})~,\\
&&G^{cd}_{3}(\tau,\tau')=-\delta_{kk'}\delta_{\alpha\beta}g_{k\alpha}(\tau,\tau')G(\tau',\tau)+\frac{V_{k\alpha}V^{*}_{k'\beta}}{\hbar^2}\nonumber\\
&&\times\iint d\tau_1 d\tau_2g_{k\alpha}(\tau,\tau_{1})g_{k'\beta}(\tau_{2},\tau')G^{dd}_{3}(\tau,\tau',\tau_{1},\tau_{2})~,\\
&&G^{cd}_{4}(\tau,\tau')=-\frac{V_{k\alpha}V_{k'\beta}}{\hbar^2}\iint d\tau_1 d\tau_2\nonumber\\
&&\times g_{k\alpha}(\tau,\tau_{1})g_{k'\beta}(\tau',\tau_{2})G^{dd}_{4}(\tau,\tau',\tau_{1},\tau_{2})~,
\end{eqnarray*}

with
\begin{eqnarray*}
G^{dd}_{1}(\tau,\tau',\tau_{1},\tau_{2})&=&i^{2}\langle T_{C}d(\tau)d(\tau')d^{\dag}(\tau_{1})d^{\dag}(\tau_{2})\rangle~,\\
G^{dd}_{2}(\tau,\tau',\tau_{1},\tau_{2})&=&i^{2}\langle T_{C}d(\tau)d^{\dag}(\tau')d(\tau_{1})d^{\dag}(\tau_{2})\rangle~,\\
G^{dd}_{3}(\tau,\tau',\tau_{1},\tau_{2})&=&i^{2}\langle T_{C}d^{\dag}(\tau)d(\tau')d(\tau_{1})d^{\dag}(\tau_{2})\rangle~,\\
G^{dd}_{4}(\tau,\tau',\tau_{1},\tau_{2})&=&i^{2}\langle T_{C}d^{\dag}(\tau)d^{\dag}(\tau')d(\tau_{1})d(\tau_{2})\rangle~.
\end{eqnarray*}

Injecting the above expressions of $G^{cd}_{i}$ in Eq.~(\ref{g1}), we get
\begin{eqnarray}\label{g2}
&&\mathcal{S}^{pq}_{\alpha\beta}(\tau,\tau')=\frac{e^{2-p-q}}{\hbar^{2}}\Bigg(\delta_{\alpha\beta}\sum_{k\in\alpha}\left(\varepsilon_{k\alpha}-\mu_{\alpha}\right)^{p+q}\nonumber\\
&&\times|V_{k\alpha}|^{2}\big[g_{k\alpha}(\tau',\tau)G(\tau,\tau')+g_{k\alpha}(\tau,\tau')G(\tau',\tau)\big]\nonumber\\
 & &+\sum_{\substack{k\in\alpha,k'\in\beta}}\left(\varepsilon_{k\alpha}-\mu_{\alpha}\right)^{p}\left(\varepsilon_{k'\beta}-\mu_{\beta}\right)^{q}\frac{|V_{k\alpha}V_{k'\beta}|^{2}}{\hbar^2}\nonumber\\
 & &\times\iint d\tau_1 d\tau_2\Big[-g_{k\alpha}(\tau_{1},\tau)g_{k'\beta}(\tau_{2},\tau')G^{dd}_{1}(\tau,\tau',\tau_{1},\tau_{2})\nonumber\\
 & &+g_{k\alpha}(\tau_{2},\tau)g_{k'\beta}(\tau',\tau_{1})G^{dd}_{2}(\tau,\tau',\tau_{1},\tau_{2})\nonumber\\
 & &  -g_{k\alpha}(\tau,\tau_{1})g_{k'\beta}(\tau_{2},\tau')G^{dd}_{3}(\tau,\tau',\tau_{1},\tau_{2})\nonumber\\
 & &-g_{k\alpha}(\tau,\tau_{1})g_{k'\beta}(\tau',\tau_{2})G^{dd}_{4}(\tau,\tau',\tau_{1},\tau_{2})\Big]\Bigg)
-\langle\hat{I}^{p}_{\alpha}\rangle\langle\hat{I}^{q}_{\beta}\rangle~,\nonumber\\
\end{eqnarray}
where $G(\tau,\tau')=-i\langle T_{C}d(\tau)d^{\dag}(\tau')\rangle$ is the one-particle Green function of the QD.

\begin{widetext}

\begin{table}[!h]
\begin{center}
\begin{tabular}{|c|c|}
  \hline
  $  \quad \mathcal{P}_0(t,t')\quad  $ & $g^{<}_{k\alpha}(t',t)G^{>}(t,t')+g^{>}_{k\alpha}(t,t')G^{<}(t',t)$\\
	\hline
	  $  \mathcal{P}_1(t,t')  $  & $-\int dt_{1}\big[G^{r}(t',t_{1})g^{<}_{k\alpha}(t_{1},t)+G^{<}(t',t_{1})g^{a}_{k\alpha}(t_{1},t)\big]\int dt_{2}\big[G^{r}(t,t_{2})g^{>}_{k'\beta}(t_{2},t')+G^{>}(t,t_{2})g^{a}_{k'\beta}(t_{2},t')\big]$\\
	\hline
	  $  \mathcal{P}_2(t,t')  $  & $G^{>}(t,t')\iint dt_{1}dt_{2}\Big[g^{r}_{k'\beta}(t',t_{1})G^{r}(t_{1},t_{2})g^{<}_{k\alpha}(t_{2},t)            $\\
	  & $            +g^{r}_{k'\beta}(t',t_{1})G^{<}(t_{1},t_{2})g^{a}_{k\alpha}(t_{2},t)+g^{<}_{k'\beta}(t',t_{1})G^{a}(t_{1},t_{2})g^{a}_{k\alpha}(t_{2},t)\Big]$\\
	\hline
	  $  \mathcal{P}_3(t,t')  $  & $G^{<}(t',t)\iint dt_{1}dt_{2}\Big[g^{r}_{k\alpha}(t,t_{1})G^{r}(t_{1},t_{2})g^{>}_{k'\beta}(t_{2},t')             $\\
	  & $             +g^{r}_{k\alpha}(t,t_{1})G^{>}(t_{1},t_{2})g^{a}_{k'\beta}(t_{2},t')+g^{>}_{k\alpha}(t,t_{1})G^{a}(t_{1},t_{2})g^{a}_{k'\beta}(t_{2},t')\Big]$\\
	\hline
	  $  \mathcal{P}_4(t,t')  $  & $-\int dt_{1}\big[g^{r}_{k\alpha}(t,t_{1})G^{>}(t_{1},t')+g^{>}_{k\alpha}(t,t_{1})G^{a}(t_{1},t')\big]$\\
&	$\times\int dt_{2}\big[g^{r}_{k'\beta}(t',t_{2})G^{<}(t_{2},t)+g^{<}_{k'\beta}(t',t_{2})G^{a}(t_{2},t)\big]$\\
	\hline
\end{tabular}
\caption{Expressions of the $\mathcal{P}_{i}(t,t')$ coefficients appearing in Eq.~(\ref{s2}).}
\label{tablePtime}
\end{center}
\end{table}

\begin{table}[!h]
\begin{center}
\begin{tabular}{|c|c|}
  \hline
	$  \quad\mathcal{P}_0(\omega)\quad  $ & $\int d\varepsilon\Big(g^{<}_{k\alpha}(\varepsilon)G^{>}(\varepsilon-\hbar\omega)+g^{>}_{k\alpha}(\varepsilon-\hbar\omega)G^{<}(\varepsilon)\Big)$\\
	\hline
	$  \mathcal{P}_1(\omega)  $  & $-\int d\varepsilon
	\Big[G^{r}(\varepsilon)g^{<}_{k\alpha}(\varepsilon)G^{r}(\varepsilon-\hbar\omega)g^{>}_{k'\beta}(\varepsilon-\hbar\omega)+G^{r}(\varepsilon)g^{<}_{k\alpha}(\varepsilon)G^{>}(\varepsilon-\hbar\omega)g^{a}_{k'\beta}(\varepsilon-\hbar\omega)$\\
		& $   +G^{<}(\varepsilon)g^{a}_{k\alpha}(\varepsilon)G^{r}(\varepsilon-\hbar\omega)g^{>}_{k'\beta}(\varepsilon-\hbar\omega)+G^{<}(\varepsilon)g^{a}_{k\alpha}(\varepsilon)G^{>}(\varepsilon-\hbar\omega)g^{a}_{k'\beta}(\varepsilon-\hbar\omega)\Big]$\\
	\hline
	$  \mathcal{P}_2(\omega)  $  & $\int d\varepsilon
	G^{>}(\varepsilon-\hbar\omega)\Big[g^{r}_{k'\beta}(\varepsilon)G^{r}(\varepsilon)g^{<}_{k\alpha}(\varepsilon)+g^{r}_{k'\beta}(\varepsilon)G^{<}(\varepsilon)g^{a}_{k\alpha}(\varepsilon)+g^{<}_{k'\beta}(\varepsilon)G^{a}(\varepsilon)g^{a}_{k\alpha}(\varepsilon)\Big]$\\
	\hline
	$  \mathcal{P}_3(\omega)  $  & $\int d\varepsilon G^{<}(\varepsilon)
	  \Big[g^{r}_{k\alpha}(\varepsilon-\hbar\omega)G^{r}(\varepsilon-\hbar\omega)g^{>}_{k'\beta}(\varepsilon-\hbar\omega)$\\
		& $    +g^{r}_{k\alpha}(\varepsilon-\hbar\omega)G^{>}(\varepsilon-\hbar\omega)g^{a}_{k'\beta}(\varepsilon-\hbar\omega) +g^{>}_{k\alpha}(\varepsilon-\hbar\omega)G^{a}(\varepsilon-\hbar\omega)g^{a}_{k'\beta}(\varepsilon-\hbar\omega)\Big]$\\
		\hline
	$  \mathcal{P}_4(\omega)  $ & $-\int d\varepsilon
\Big[g^{r}_{k\alpha}(\varepsilon-\hbar\omega)G^{>}(\varepsilon-\hbar\omega)+g^{>}_{k\alpha}(\varepsilon-\hbar\omega)G^{a}(\varepsilon-\hbar\omega)\Big]
\Big[g^{r}_{k'\beta}(\varepsilon)G^{<}(\varepsilon)+g^{<}_{k'\beta}(\varepsilon)G^{a}(\varepsilon)\Big]$\\
		\hline
\end{tabular}
\caption{Expressions of the $\mathcal{P}_{i}(\omega)$ coefficients appearing in Eq.~(\ref{s3}).}
\label{tablePomega}
\end{center}
\end{table}

\begin{table}[!h]
\begin{center}
\begin{tabular}{|c|c|}
  \hline
	$A_{\alpha\beta}(\varepsilon,\omega)$ & $-\frac{1}{2}\big[f^h_{\alpha}(\varepsilon-\hbar\omega)+f^h_{\bar{\alpha}}(\varepsilon-\hbar\omega)\big]\big([f^e_{\alpha}(\varepsilon)-f^e_{\beta}(\varepsilon)]t(\varepsilon)\mathcal{T}(\varepsilon-\hbar\omega)-\frac{1}{2}[f^e_{\bar{\beta}}(\varepsilon)-3f^e_{\beta}(\varepsilon)]\mathcal{T}(\varepsilon)\mathcal{T}(\varepsilon-\hbar\omega)\big)$\\
	\hline
	$B_{\alpha\beta}(\varepsilon,\omega)$ & $f^e_{\alpha}(\varepsilon)f^h_{\beta}(\varepsilon-\hbar\omega)t(\varepsilon)t(\varepsilon-\hbar\omega)-\frac{1}{2}f^e_{\alpha}(\varepsilon)\big[f^h_{\alpha}(\varepsilon-\hbar\omega)+f^h_{\bar{\alpha}}(\varepsilon-\hbar\omega)\big]t(\varepsilon)\mathcal{T}(\varepsilon-\hbar\omega)$\\
	&$-\frac{1}{2}f^h_{\beta}(\varepsilon-\hbar\omega)\big[f^e_{\alpha}(\varepsilon)+f^e_{\bar{\alpha}}(\varepsilon)\big]\mathcal{T}(\varepsilon)t(\varepsilon-\hbar\omega)+\frac{1}{4}\big[f^e_{\alpha}(\varepsilon)+f^e_{\bar{\alpha}}(\varepsilon)\big]\big[f^h_{\alpha}(\varepsilon-\hbar\omega)+f^h_{\bar{\alpha}}(\varepsilon-\hbar\omega)\big]\mathcal{T}(\varepsilon)\mathcal{T}(\varepsilon-\hbar\omega)$\\
	\hline
	$C_{\alpha\beta}(\varepsilon,\omega)$ & $-\frac{1}{2}f^h_{\beta}(\varepsilon-\hbar\omega)\big[f^e_{\beta}(\varepsilon)+f^e_{\bar{\beta}}(\varepsilon)\big]\mathcal{T}(\varepsilon)t(\varepsilon-\hbar\omega)-\frac{1}{2}f^h_{\alpha}(\varepsilon-\hbar\omega)\big[f^e_{\beta}(\varepsilon)+f^e_{\bar{\beta}}(\varepsilon)\big]\mathcal{T}(\varepsilon)t^{*}(\varepsilon-\hbar\omega)      $\\
	&$   +\frac{1}{4}\big[f^e_{\beta}(\varepsilon)+f^e_{\bar{\beta}}(\varepsilon)\big]\big[f^h_{\beta}(\varepsilon-\hbar\omega)+f^h_{\bar{\beta}}(\varepsilon-\hbar\omega)\big]\mathcal{T}(\varepsilon)\mathcal{T}(\varepsilon-\hbar\omega)$\\
	\hline
  	$D_{\alpha}(\varepsilon,\omega)$ & $f^e_{\alpha}(\varepsilon)\big[f^h_{\alpha}(\varepsilon-\hbar\omega)+f^h_{\bar{\alpha}}(\varepsilon-\hbar\omega)\big]\mathcal{T}(\varepsilon-\hbar\omega)$\\
	\hline
	$E_{\alpha}(\varepsilon,\omega)$ & $\big[f^e_{\alpha}(\varepsilon)+f^e_{\bar{\alpha}}(\varepsilon)\big]f^h_{\alpha}(\varepsilon-\hbar\omega)\mathcal{T}(\varepsilon)$\\
	\hline
\end{tabular}
\end{center}
\caption{Expressions of the coefficients appearing in Eq.~(\ref{s5}).}
\label{tablecoeff}
\end{table}

\end{widetext}

\subsubsection{Evaluation of the two-particle Green functions $G^{dd}_i$ using decoupling procedure}

In the absence of Coulomb interactions, we can express fully the two-particle Green functions of the QD, $G^{dd}_i(\tau,\tau',\tau_{1},\tau_{2})$, in terms of the one-particle Green function of the QD, $G(\tau,\tau')$, through
\begin{eqnarray*}
G^{dd}_{1}(\tau,\tau',\tau_{1},\tau_{2})&=&G(\tau,\tau_{2})G(\tau',\tau_{1})-G(\tau,\tau_{1})G(\tau',\tau_{2})~,\\
G^{dd}_{2}(\tau,\tau',\tau_{1},\tau_{2})&=&G(\tau,\tau')G(\tau_{1},\tau_{2})-G(\tau,\tau_{2})G(\tau_{1},\tau')~,\\
G^{dd}_{3}(\tau,\tau',\tau_{1},\tau_{2})&=&G(\tau_{1},\tau)G(\tau',\tau_{2})-G(\tau',\tau)G(\tau_{1},\tau_{2})~,\\
G^{dd}_{4}(\tau,\tau',\tau_{1},\tau_{2})&=&G(\tau_{2},\tau)G(\tau_{1},\tau')-G(\tau_{1},\tau)G(\tau_{2},\tau')~.
\end{eqnarray*}
Injecting these four expressions in Eq.~(\ref{g2}), we obtain a result which can be separated into two parts, a \textit{connected} part and a \textit{disconnected} part \cite{Haug2007}
\begin{eqnarray}
\mathcal{S}^{pq}_{\alpha\beta}(\tau,\tau')&=&\mathcal{S}^{pq}_{\substack{\alpha\beta,disc}}(\tau,\tau')+\mathcal{S}^{pq}_{\substack{\alpha\beta,conn}}(\tau,\tau')-\langle\hat{I}^{p}_{\alpha}\rangle\langle\hat{I}^{q}_{\beta}\rangle~,\nonumber\\
\end{eqnarray}
with 
 \begin{eqnarray}
&&\mathcal{S}^{pq}_{\substack{\alpha\beta,disc}}(\tau,\tau')=\frac{e^{2-p-q}}{\hbar^{2}}\nonumber\\
&&\times\sum_{\substack{k\in\alpha,k'\in\beta}}\left(\varepsilon_{k\alpha}-\mu_{\alpha}\right)^{p}\left(\varepsilon_{k'\beta}-\mu_{\beta}\right)^{q}\frac{|V_{k\alpha}V_{k'\beta}|^{2}}{\hbar^{2}}\nonumber\\
&& \times\bigg[\int d\tau_{1}g_{k\alpha}(\tau_{1},\tau)G(\tau,\tau_{1})\int d\tau_{2}g_{k'\beta}(\tau_{2},\tau')G(\tau',\tau_{2})\nonumber\\
&&
-\int d\tau_{2}g_{k\alpha}(\tau_{2},\tau)G(\tau,\tau_{2})\int d\tau_{1}g_{k'\beta}(\tau',\tau_{1})G(\tau_{1},\tau')\nonumber\\
& &-\int d\tau_{1}g_{k\alpha}(\tau,\tau_{1})G(\tau_{1},\tau)\int d\tau_{2}g_{k'\beta}(\tau_{2},\tau')G(\tau',\tau_{2})\nonumber\\
&&
+\int d\tau_{1}g_{k\alpha}(\tau,\tau_{1})G(\tau_{1},\tau)\int d\tau_{2}g_{k'\beta}(\tau',\tau_{2})G(\tau_{2},\tau')\bigg]~,\nonumber\\
\end{eqnarray}
and
\begin{eqnarray}\label{s4}
&&\mathcal{S}^{pq}_{\substack{\alpha\beta,conn}}(\tau,\tau')=\frac{e^{2-p-q}}{\hbar^{2}}\bigg[\delta_{\alpha\beta}\sum_{k\in\alpha}\left(\varepsilon_{k\alpha}-\mu_{\alpha}\right)^{p+q}|V_{k\alpha}|^{2}\nonumber\\
&&\times\big[g_{k\alpha}(\tau',\tau)G(\tau,\tau')+g_{k\alpha}(\tau,\tau')G(\tau',\tau)\big]\nonumber\\
& &+\sum_{\substack{k\in\alpha,k'\in\beta}}\left(\varepsilon_{k\alpha}-\mu_{\alpha}\right)^{p}\left(\varepsilon_{k'\beta}-\mu_{\beta}\right)^{q}\frac{|V_{k\alpha}V_{k'\beta}|^{2}}{\hbar^{2}}\nonumber\\
& &\times\iint d\tau_1 d\tau_2\Big[-g_{k\alpha}(\tau_{1},\tau)g_{k'\beta}(\tau_{2},\tau')G(\tau,\tau_{2})G(\tau',\tau_{1})\nonumber\\
&&+g_{k\alpha}(\tau_{2},\tau)g_{k'\beta}(\tau',\tau_{1})G(\tau,\tau')G(\tau_{1},\tau_{2})\nonumber\\
& &  +g_{k\alpha}(\tau,\tau_{1})g_{k'\beta}(\tau_{2},\tau')G(\tau',\tau)G(\tau_{1},\tau_{2})\nonumber\\
&&-g_{k\alpha}(\tau,\tau_{1})g_{k'\beta}(\tau',\tau_{2})G(\tau_{2},\tau)G(\tau_{1},\tau')\Big]\bigg]~.
\end{eqnarray}

The disconnected part can be calculated directly. We obtain $\mathcal{S}^{pq}_{\substack{\alpha\beta,disc}}(\tau,\tau')=\langle\hat{I}^{p}_{\alpha}\rangle\langle\hat{I}^{q}_{\beta}\rangle$,
thus finally $\mathcal{S}^{pq}_{\alpha\beta}(\tau,\tau')=\mathcal{S}^{pq}_{\substack{\alpha\beta,conn}}(\tau,\tau')$.

\subsubsection{Analytic continuation of the connected part}

We perform now the analytic continuation of Eq.~(\ref{s4}) to get its $\tau>\tau'$ component
\begin{eqnarray*}
&&\mathcal{S}^{pq}_{\alpha\beta}(t,t')=
\frac{e^{2-p-q}}{\hbar^{2}}\Bigg[\delta_{\alpha\beta}\sum_{k\in\alpha}\left(\varepsilon_{k\alpha}-\mu_{\alpha}\right)^{p+q}|V_{k\alpha}|^{2}\nonumber\\
& &\times\underbrace{\big[g_{k\alpha}(\tau',\tau)G(\tau,\tau')+g_{k\alpha}(\tau,\tau')G(\tau',\tau)\big]_{\tau>\tau'}}_{\mathcal{P}_{0}(t,t')}\\
& &+\sum_{\substack{k\in\alpha,k'\in\beta}}\left(\varepsilon_{k\alpha}-\mu_{\alpha}\right)^{p}\left(\varepsilon_{k'\beta}-\mu_{\beta}\right)^{q}\frac{|V_{k\alpha}V_{k'\beta}|^{2}}{\hbar^{2}}\\
& &\times\bigg(-\underbrace{\big[\iint d\tau_{1}d\tau_{2}G(\tau',\tau_{1})g_{k\alpha}(\tau_{1},\tau)G(\tau,\tau_{2})g_{k'\beta}(\tau_{2},\tau')\big]_{\tau>\tau'}}_{\mathcal{P}_{1}(t,t')}\nonumber\\
& &
+\underbrace{\big[G(\tau,\tau')\iint d\tau_{1}d\tau_{2}g_{k'\beta}(\tau',\tau_{1})G(\tau_{1},\tau_{2})g_{k\alpha}(\tau_{2},\tau)\big]_{\tau>\tau'}}_{\mathcal{P}_{2}(t,t')}\\
& &  +\underbrace{\big[G(\tau',\tau)\iint d\tau_{1}d\tau_{2}g_{k\alpha}(\tau,\tau_{1})G(\tau_{1},\tau_{2})g_{k'\beta}(\tau_{2},\tau')\big]_{\tau>\tau'}}_{\mathcal{P}_{3}(t,t')}\nonumber\\
& &
-\underbrace{\big[\iint d\tau_{1}d\tau_{2}g_{k\alpha}(\tau,\tau_{1})G(\tau_{1},\tau')g_{k'\beta}(\tau',\tau_{2})G(\tau_{2},\tau)\big]_{\tau>\tau'}}_{\mathcal{P}_{4}(t,t')}\bigg)\Bigg]~.
\end{eqnarray*}
The five $\mathcal{P}_{i}(t,t')$ contributions are computed using analytic continuation rules \cite{Haug2007}. Their expressions are given in Table~\ref{tablePtime}. Using these notations, the noise reads as
\begin{eqnarray}\label{s2}
&&\mathcal{S}^{pq}_{\alpha\beta}(t,t')=\frac{e^{2-p-q}}{\hbar^2}\Bigg[\delta_{\alpha\beta}\sum_{k\in\alpha}\left(\varepsilon_{k\alpha}-\mu_{\alpha}\right)^{p+q}|V_{k\alpha}|^{2}\mathcal{P}_{0}(t,t')\nonumber\\
&&+\sum_{\substack{k\in\alpha\\k'\in\beta}}\left(\varepsilon_{k\alpha}-\mu_{\alpha}\right)^{p}\left(\varepsilon_{k'\beta}-\mu_{\beta}\right)^{q}\frac{|V_{k\alpha}V_{k'\beta}|^{2}}{\hbar^{2}}\sum^{4}_{i=1}\mathcal{P}_{i}(t,t')\Bigg]~.\nonumber\\
\end{eqnarray}

\subsection{Finite-frequency non-symmetrized noise $\mathcal{S}^{pq}_{\alpha\beta}(\omega)$}

\subsubsection{Fourier transform of $\mathcal{S}^{pq}_{\alpha\beta}(t,t')$ and exact result for $\mathcal{S}^{pq}_{\alpha\beta}(\omega)$}

Performing the Fourier transform of Eq.~(\ref{s2}) and using the fact that in the stationary case the Green functions depend on the difference of their time arguments only, we obtain:
\begin{eqnarray}\label{s3}
&&\mathcal{S}^{pq}_{\alpha\beta}(\omega)=\frac{e^{2-p-q}}{\hbar}\Bigg[\delta_{\alpha\beta}\sum_{k\in\alpha}\left(\varepsilon_{k\alpha}-\mu_{\alpha}\right)^{p+q}|V_{k\alpha}|^{2}\mathcal{P}_{0}(\omega)\nonumber\\
&&+\sum_{\substack{k\in\alpha\\k'\in\beta}}\left(\varepsilon_{k\alpha}-\mu_{\alpha}\right)^{p}\left(\varepsilon_{k'\beta}-\mu_{\beta}\right)^{q}|V_{k\alpha}V_{k'\beta}|^{2}\sum^{4}_{i=1}\mathcal{P}_{i}(\omega)\Bigg]~,\nonumber\\
\end{eqnarray}
where the expressions of $\mathcal{P}_{i}(\omega)$ are given in Table~\ref{tablePomega}. Using the expressions of the bare Green functions of the reservoirs, $g^{>}_{k\alpha}(\varepsilon)$ and $g^{<}_{k\alpha}(\varepsilon)$, in terms of the Fermi-Dirac distribution function for electrons, $f^e_{\alpha}(\varepsilon)$, and Fermi-Dirac distribution function for holes, $f^h_{\alpha}(\varepsilon)=1-f^e_{\alpha}(\varepsilon)$,
\begin{eqnarray}
&&g^{<}_{k\alpha}(\varepsilon)=2\pi if^e_{\alpha}(\varepsilon)\delta(\varepsilon-\varepsilon_{k\alpha})~,\\
&&g^{>}_{k\alpha}(\varepsilon)=-2\pi if^h_{\alpha}(\varepsilon)\delta(\varepsilon-\varepsilon_{k\alpha})~,
\end{eqnarray}
we can rewrite Eq.~(\ref{s3}) under the form
\begin{eqnarray}\label{s1}
&&\mathcal{S}^{pq}_{\alpha\beta}(\omega)=\frac{e^{2-p-q}}{h}\int d\varepsilon 
\Bigg[\delta_{\alpha\beta}i F^{p+q}_{\alpha}(\varepsilon)f^e_{\alpha}(\varepsilon)\nonumber\\
			& &\times\Big[G^{<}(\varepsilon-\hbar\omega)+G^{r}(\varepsilon-\hbar\omega)-G^{a}(\varepsilon-\hbar\omega)\Big]\nonumber\\
			& &-\delta_{\alpha\beta}i F^{p+q}_{\alpha}(\varepsilon-\hbar\omega)f^h_{\alpha}(\varepsilon-\hbar\omega)G^{<}(\varepsilon)\nonumber\\
			& &
			-G^{r}(\varepsilon)G^{r}(\varepsilon-\hbar\omega)f^e_{\alpha}(\varepsilon)f^h_{\beta}(\varepsilon-\hbar\omega)F^{p}_{\alpha}(\varepsilon)F^{q}_{\beta}(\varepsilon-\hbar\omega)\nonumber\\
			& &-G^{a}(\varepsilon)G^{a}(\varepsilon-\hbar\omega)f^e_{\beta}(\varepsilon)f^h_{\alpha}(\varepsilon-\hbar\omega)F^{p}_{\alpha}(\varepsilon-\hbar\omega)F^{q}_{\beta}(\varepsilon)\nonumber\\
			& &
			+iG^{r}(\varepsilon)G^{>}(\varepsilon-\hbar\omega) f^e_{\alpha}(\varepsilon)F^{p}_{\alpha}(\varepsilon)H^{q*}_{\beta}(\varepsilon)\nonumber\\
			& &+iG^{<}(\varepsilon)G^{r}(\varepsilon-\hbar\omega)f^h_{\beta}(\varepsilon-\hbar\omega)F^{q}_{\beta}(\varepsilon-\hbar\omega)H^{p}_{\alpha}(\varepsilon)\nonumber\\
			& &
			+iG^{a}(\varepsilon)G^{>}(\varepsilon-\hbar\omega) f^e_{\beta}(\varepsilon)F^{q}_{\beta}(\varepsilon)H^{p}_{\alpha}(\varepsilon)\nonumber\\
			& &+iG^{<}(\varepsilon)G^{a}(\varepsilon-\hbar\omega)f^h_{\alpha}(\varepsilon-\hbar\omega)F^{p}_{\alpha}(\varepsilon-\hbar\omega)H^{q*}_{\beta}(\varepsilon)\nonumber\\
			& &
			+G^{<}(\varepsilon)G^{>}(\varepsilon-\hbar\omega)H^{p}_{\alpha}(\varepsilon,p)H^{q*}_{\beta}(\varepsilon)\Bigg]~,
\end{eqnarray}
where we have introduced the two following functions
\begin{eqnarray}
F^p_{\alpha}(\varepsilon)&=&2\pi\sum_{k\in\alpha}\left(\varepsilon_{k\alpha}-\mu_{\alpha}\right)^{p}|V_{k\alpha}|^{2}\delta(\varepsilon-\varepsilon_{k\alpha})\nonumber\\
&=&
\left(\varepsilon-\mu_{\alpha}\right)^{p}\underbrace{2\pi |V_\alpha(\varepsilon)|^{2}\rho_\alpha(\varepsilon)}_
{=\Gamma_{\alpha}(\varepsilon)}~,
\end{eqnarray}
and
\begin{eqnarray}
H^p_{\alpha}(\varepsilon)&=&\sum_{k\in\alpha}\left(\varepsilon_{k\alpha}-\mu_{\alpha}\right)^{p}|V_{k\alpha}|^{2}\big[g^{a}_{k\alpha}(\varepsilon)-g^{r}_{k\alpha}(\varepsilon-\hbar\omega)\big]~,\nonumber\\
\end{eqnarray}
with $\rho_\alpha(\varepsilon)$ the density of states associated to the reservoirs $\alpha$, and $\Gamma_\alpha=2\pi |V_\alpha(\varepsilon)|^{2}\rho_\alpha(\varepsilon)$, the coupling strength between the QD and the reservoir $\alpha$. Note that Eq.~(\ref{s1}), given $\mathcal{S}^{pq}_{\alpha\beta}(\omega)$, has been obtained without making any approximation at this stage: it is the exact result for a non-interacting QD.

\subsubsection{$\mathcal{S}^{pq}_{\alpha\beta}(\omega)$ for symmetrical barriers in the wide-band limit}

To continue further the calculation, we make two simplifying assumptions:
(i) wide-band limit, i.e., we are working on an interval of energy in which the density of states is constant, $\rho_{\alpha}(\varepsilon)=cst$, and we assume as well that $V_{\alpha}(\varepsilon)$ is energy independent, consequently, we have $\Gamma_\alpha(\varepsilon)=2\pi |V_\alpha(\varepsilon)|^2 \rho_{\alpha}(\varepsilon)\equiv\Gamma_\alpha$, and (ii) symmetrical barriers, i.e., we assume that the left and right barriers are symmetrical ($\Gamma_{L}=\Gamma_{R}\equiv\Gamma$). In that case, we have the remarkable relation \cite{Zamoum2016}: $t(\varepsilon)+t^{*}(\varepsilon)=2\mathcal{T}(\varepsilon)$, with $t(\varepsilon)=i\Gamma G^{r}(\varepsilon)$, the transmission amplitude and $\mathcal{T}(\varepsilon)$, the transmission coefficient.
Within these two simplifying assumptions, we have
\begin{eqnarray*}
F^p_{\alpha}(\varepsilon)&=&\Gamma(\varepsilon-\mu_{\alpha})^{p}~,
\end{eqnarray*}
and
\begin{eqnarray*}
H^p_{\alpha}(\varepsilon)&=&\frac{i\Gamma}{2}\left[(\varepsilon-\mu_{\alpha})^{p}+(\varepsilon-\hbar\omega-\mu_{\alpha})^{p}\right]~.
\end{eqnarray*}
Injecting these two last expressions in Eq.~(\ref{s1}), rearranging the terms, and using the relations \cite{Mahan2000,Haug2007}
\begin{eqnarray*}
&&G^{>}(\varepsilon)-G^{<}(\varepsilon)=G^{r}(\varepsilon)-G^{a}(\varepsilon)~,\\
&&G^{<}(\varepsilon)=i\Gamma G^{r}(\varepsilon)G^{a}(\varepsilon)\big[f^e_{\alpha}(\varepsilon)+f^e_{\bar{\alpha}}(\varepsilon)\big]~,\\
&&G^{r}(\varepsilon)-G^{a}(\varepsilon)=-2i\Gamma G^{r}(\varepsilon)G^{a}(\varepsilon)~,\\
&&\mathcal{T}(\varepsilon)=\Gamma^2G^{r}(\varepsilon)G^{a}(\varepsilon)~.
\end{eqnarray*}
We finally get 
\begin{eqnarray}\label{s5}
    &&\mathcal{S}^{pq}_{\alpha\beta}(\omega)=\frac{e^{2-p-q}}{h}\int d\varepsilon
		\bigg[\delta_{\alpha\beta}\left(\varepsilon-\mu_{\alpha}\right)^{p+q}D_{\alpha}(\varepsilon,\omega)\nonumber\\
		& & +\delta_{\alpha\beta}\left(\varepsilon-\hbar\omega-\mu_{\alpha}\right)^{p+q}E_{\alpha}(\varepsilon,\omega)\nonumber\\
		& & +\left(\varepsilon-\mu_{\alpha}\right)^{p}\left(\varepsilon-\mu_{\beta}\right)^{q}A_{\alpha\beta}(\varepsilon,\omega)
		\nonumber\\
		& & +\left(\varepsilon-\mu_{\alpha}\right)^{p}\left(\varepsilon-\hbar\omega-\mu_{\beta}\right)^{q}B_{\alpha\beta}(\varepsilon,\omega)\nonumber\\
		& & +\left(\varepsilon-\hbar\omega-\mu_{\alpha}\right)^{p}\left(\varepsilon-\mu_{\beta}\right)^{q}B^{*}_{\beta\alpha}(\varepsilon,\omega)\nonumber\\
		& & +\left(\varepsilon-\hbar\omega-\mu_{\alpha}\right)^{p}\left(\varepsilon-\hbar\omega-\mu_{\beta}\right)^{q}C_{\alpha\beta}(\varepsilon,\omega)\bigg]~,
\end{eqnarray}
where the coefficients $A_{\alpha}(\varepsilon,\omega)$, $B_{\alpha}(\varepsilon,\omega)$,  $C_{\alpha}(\varepsilon,\omega)$, $D_{\alpha}(\varepsilon,\omega)$, and $E_{\alpha}(\varepsilon,\omega)$ are given in Table~\ref{tablecoeff}. Equation~(\ref{s5}) leads to the Eqs.~(\ref{exp_noise})-(\ref{C}) once we define
\begin{eqnarray*}
\mathcal{A}_{\alpha\beta}(\varepsilon,\omega)&=&A_{\alpha\beta}(\varepsilon,\omega)+\delta_{\alpha\beta}D_{\alpha}(\varepsilon,\omega)~,\\
\mathcal{B}_{\alpha\beta}(\varepsilon,\omega)&=&B_{\alpha\beta}(\varepsilon,\omega)~,\\
\mathcal{C}_{\alpha\beta}(\varepsilon,\omega)&=&C_{\alpha\beta}(\varepsilon,\omega)+\delta_{\alpha\beta}E_{\alpha}(\varepsilon,\omega)~,\\
f^{e,h}_M(\varepsilon)&=&\frac{1}{2}\big[f^{e,h}_\alpha(\varepsilon)+f^{e,h}_{\bar\alpha}(\varepsilon)\big]~,
\end{eqnarray*}
where $\bar{\alpha}=R$ when $\alpha=L$, and $\bar{\alpha}=L$ when $\alpha=R$.

%
%
%
%
%
%

\section{Noise spectrum for $t$ and $\mathcal{T}$ independent of energy}
\label{appendixB}

In case of independent energy transmission amplitude and coefficient, Eqs.~(\ref{exp_noise})-(\ref{C}) reduce to
\begin{eqnarray}\label{S00}
\mathcal{S}^{00}_{\alpha\beta}(\omega)&=&\frac{e^{2}}{h}\sum_{\gamma\delta}\int_{-\infty}^{\infty} d\varepsilon \mathcal{M}_{\alpha\beta}^{\gamma\delta}f^e_\gamma(\varepsilon)f^h_\delta(\varepsilon-\hbar\omega)~,\nonumber\\
\end{eqnarray}
\begin{eqnarray}\label{S01}
\mathcal{S}^{01}_{\alpha\beta}(\omega)&=&\frac{e}{h}\sum_{\gamma\delta}\int_{-\infty}^{\infty} d\varepsilon 
\big[(\varepsilon-\mu_\beta)\mathcal{M}_{\alpha\beta}^{\gamma\delta}-\hbar\omega\mathcal{N}_{\alpha\beta}^{\gamma\delta}\big]\nonumber\\
&&\times f^e_\gamma(\varepsilon)f^h_\delta(\varepsilon-\hbar\omega)~,
\end{eqnarray}
\begin{eqnarray}\label{S10}
\mathcal{S}^{10}_{\alpha\beta}(\omega)&=&\frac{e}{h}\sum_{\gamma\delta}\int_{-\infty}^{\infty} d\varepsilon 
\big[(\varepsilon-\mu_\alpha)\mathcal{M}_{\alpha\beta}^{\gamma\delta}
-\hbar\omega(\mathcal{N}_{\beta\alpha}^{\gamma\delta})^*\big]\nonumber\\
&&\times f^e_\gamma(\varepsilon)f^h_\delta(\varepsilon-\hbar\omega)~,
\end{eqnarray}
and
\begin{eqnarray}\label{S11}
\mathcal{S}^{11}_{\alpha\beta}(\omega)&=&\frac{1}{h}\sum_{\gamma\delta}\int_{-\infty}^{\infty} d\varepsilon 
\big[(\varepsilon-\mu_\alpha)(\varepsilon-\mu_\beta)\mathcal{M}_{\alpha\beta}^{\gamma\delta}\nonumber\\
&&-\hbar\omega(\varepsilon-\mu_\alpha)\mathcal{N}_{\alpha\beta}^{\gamma\delta}-\hbar\omega(\varepsilon-\mu_\beta)(\mathcal{N}_{\beta\alpha}^{\gamma\delta})^*\nonumber\\
&&+\hbar^2\omega^2\mathcal{O}_{\alpha\beta}^{\gamma\delta}\big]f^e_\gamma(\varepsilon)f^h_\delta(\varepsilon-\hbar\omega)~,
\end{eqnarray}
with the coefficients $\mathcal{M}_{\alpha\beta}^{\gamma\delta}$, $\mathcal{N}_{\alpha\beta}^{\gamma\delta}$, and $\mathcal{O}_{\alpha\beta}^{\gamma\delta}$ given in Tables~\ref{tableM}-\ref{tableO}, where $\mathcal{Z}=[\mathcal{T}(1-\mathcal{T})]^{1/2}$ is the imaginary part of $t$, $\mathcal{T}$ being the real part of $t$. These real and imaginary parts are extracted from the two relations: $tt^*=\mathcal{T}$ and $t+t^*=2\mathcal{T}$.

\begin{widetext}

\begin{table}[!h]
\begin{center}
\begin{tabular}{|c||c|c|c|c|}
\hline
$\mathcal{M}_{\alpha\beta}^{\gamma\delta}$& $\gamma=\delta=L$& $\gamma=\delta=R$&$\gamma=L$, $\delta=R$&$\gamma=R$, $\delta=L$\\ \hline\hline
$\alpha=\beta=L$&  $\mathcal{T}^2$& $\mathcal{T}^2$ & $\mathcal{T}(1-\mathcal{T})$ &  $\mathcal{T}(1-\mathcal{T})$\\
 \hline
$\alpha=\beta=R$&  $\mathcal{T}^2$& $\mathcal{T}^2$ & $\mathcal{T}(1-\mathcal{T})$ &  $\mathcal{T}(1-\mathcal{T})$\\
 \hline
$\alpha=L$, $\beta=R$&  $-\mathcal{T}^2$& $-\mathcal{T}^2$ & $-\mathcal{T}(1-\mathcal{T})$ &  $-\mathcal{T}(1-\mathcal{T})$\\
 \hline
$\alpha=R$,$\beta=L$&  $-\mathcal{T}^2$& $-\mathcal{T}^2$ & $-\mathcal{T}(1-\mathcal{T})$ &  $-\mathcal{T}(1-\mathcal{T})$\\
 \hline
\end{tabular}
\end{center}
\caption{Expressions of the coefficients $\mathcal{M}_{\alpha\beta}^{\gamma\delta}$ appearing in Eqs.~(\ref{S00})-(\ref{S11}).}
\label{tableM}
\end{table}

\begin{table}[!h]
\begin{center}
\begin{tabular}{|c||c|c|c|c|}
\hline
$\mathcal{N}_{\alpha\beta}^{\gamma\delta}$& $\gamma=\delta=L$& $\gamma=\delta=R$&$\gamma=L$, $\delta=R$&$\gamma=R$, $\delta=L$\\ \hline\hline
$\alpha=\beta=L$&$\frac{\mathcal{T}^2}{2}+i\mathcal{ZT}$  &$\frac{\mathcal{T}^2}{2}$&$-\frac{i\mathcal{ZT}}{2}$&$\mathcal{T}(1-\mathcal{T})-\frac{i\mathcal{ZT}}{2}$\\
 \hline
$\alpha=\beta=R$&$\frac{\mathcal{T}^2}{2}$  &$\frac{\mathcal{T}^2}{2}+i\mathcal{ZT}$&$\mathcal{T}(1-\mathcal{T})-\frac{i\mathcal{ZT}}{2}$ &$-\frac{i\mathcal{ZT}}{2}$ \\
 \hline
$\alpha=L$, $\beta=R$&$-\frac{\mathcal{T}^2}{2}$ &$-\frac{\mathcal{T}^2}{2}-i\mathcal{ZT}$&$-\mathcal{T}(1-\mathcal{T})+\frac{i\mathcal{ZT}}{2}$&$\frac{i\mathcal{ZT}}{2}$\\
 \hline
$\alpha=R$,$\beta=L$& $-\frac{\mathcal{T}^2}{2}-i\mathcal{ZT}$&$-\frac{\mathcal{T}^2}{2}$&$\frac{i\mathcal{ZT}}{2}$&$-\mathcal{T}(1-\mathcal{T})+\frac{i\mathcal{ZT}}{2}$\\
 \hline
\end{tabular}
\end{center}
\caption{Expressions of the coefficients $\mathcal{N}_{\alpha\beta}^{\gamma\delta}$ appearing in Eqs.~(\ref{S01})-(\ref{S11}).}
\label{tableN}
\end{table}

\begin{table}[!h]
\begin{center}
\begin{tabular}{|c||c|c|c|c|}
\hline
$\mathcal{O}_{\alpha\beta}^{\gamma\delta}$& $\gamma=\delta=L$& $\gamma=\delta=R$&$\gamma=L$, $\delta=R$&$\gamma=R$, $\delta=L$\\ \hline\hline
$\alpha=\beta=L$&$\frac{\mathcal{T}^2}{4}+\mathcal{T}(1-\mathcal{T})$&$\frac{\mathcal{T}^2}{4}$&$\frac{\mathcal{T}^2}{4}$&$\frac{\mathcal{T}^2}{4}+\mathcal{T}(1-\mathcal{T})$\\
 \hline
$\alpha=\beta=R$&$\frac{\mathcal{T}^2}{4}$&$\frac{\mathcal{T}^2}{4}+\mathcal{T}(1-\mathcal{T})$&$\frac{\mathcal{T}^2}{4}+\mathcal{T}(1-\mathcal{T})$&$\frac{\mathcal{T}^2}{4}$\\
 \hline
$\alpha=L$, $\beta=R$&$-\frac{\mathcal{T}^2}{4}+\frac{i\mathcal{ZT}}{2}$&$-\frac{\mathcal{T}^2}{4}-\frac{i\mathcal{ZT}}{2}$&$-\frac{\mathcal{T}^2}{4}-\frac{i\mathcal{ZT}}{2}$&$-\frac{\mathcal{T}^2}{4}+\frac{i\mathcal{ZT}}{2}$\\
 \hline
$\alpha=R$,$\beta=L$&$-\frac{\mathcal{T}^2}{4}-\frac{i\mathcal{ZT}}{2}$&$-\frac{\mathcal{T}^2}{4}+\frac{i\mathcal{ZT}}{2}$&$-\frac{\mathcal{T}^2}{4}+\frac{i\mathcal{ZT}}{2}$&$-\frac{\mathcal{T}^2}{4}-\frac{i\mathcal{ZT}}{2}$\\
 \hline
\end{tabular}
\end{center}
\caption{Expressions of the coefficients $\mathcal{O}_{\alpha\beta}^{\gamma\delta}$ appearing in Eq.~(\ref{S11}).}
\label{tableO}
\end{table}

\end{widetext}

\subsection{Preliminary calculations}

In the following sections, we will meet the integral
\begin{eqnarray}
 I^{(n)}_{\gamma\delta}=\int_{-\infty}^\infty \varepsilon^n d\varepsilon f_\gamma^e(\varepsilon)f_\delta^h(\varepsilon-\hbar\omega)~,
\end{eqnarray}
with $n$ = 0, 1 or 2. Here, we calculate this integral considering the isothermal case, $T_L=T_R\equiv T$, thus
\pagebreak
\begin{eqnarray}
&& I^{(n)}_{\gamma\delta}
=N(\hbar\omega+\mu_\delta-\mu_\gamma)\nonumber\\
&&\times\int_{-\infty}^\infty \varepsilon^n d\varepsilon \frac{\sinh\left(\frac{\hbar\omega+\mu_\delta-\mu_\gamma}{2k_BT}\right)}{2\cosh\left(\frac{\varepsilon-\mu_\gamma}{2k_BT}\right)\cosh\left(\frac{\varepsilon-\mu_\delta-\hbar\omega}{2k_BT}\right)}~,
\end{eqnarray}
where we have introduced the Bose-Einstein distribution function: $N(\hbar\omega)=[\exp(\hbar\omega/k_BT)-1]^{-1}$. Using the identity: $\sinh(a-b)/[\cosh(a)\cosh(b)]=\tanh(a)-\tanh(b)$, we end up with
\begin{eqnarray}
 &&I^{(n)}_{\gamma\delta}
=\frac{N(\hbar\omega+\mu_\delta-\mu_\gamma)}{2}\nonumber\\
&&\times\int_{-\infty}^\infty \varepsilon^n d\varepsilon \left[\tanh\left(\frac{\varepsilon-\mu_\gamma}{2k_BT}\right)-\tanh\left(\frac{\varepsilon-\mu_\delta-\hbar\omega}{2k_BT}\right)\right]~.\nonumber\\
\end{eqnarray}
To go further, we perform a Taylor expansion up to the third order with $x=\omega$, $\mu_\gamma$ or $\mu_\delta$. It leads to
\begin{eqnarray}
I^{(n)}_{\gamma\delta}
&=&\frac{N(\hbar\omega+\mu_\delta-\mu_\gamma)}{2}\Bigg[\frac{\mu_\gamma-\mu_\delta-\hbar\omega}{2k_BT}L^{(n)}_1 \nonumber\\
&&+\frac{\mu_\gamma^2-(\mu_\delta+\hbar\omega)^2}{4k_B^2T^2}L^{(n)}_2  +\frac{\mu_\gamma^3-(\mu_\delta+\hbar\omega)^3}{24k_B^3T^3}L^{(n)}_3\Bigg]~,\nonumber\\
\end{eqnarray}
with $L^{(n)}_1=\int_{-\infty}^\infty \varepsilon^n d\varepsilon[\tanh^2(\varepsilon/2k_BT)-1]$, $L^{(n)}_2=\int_{-\infty}^\infty \varepsilon^n d\varepsilon[\tanh^3(\varepsilon/2k_BT)-\tanh(\varepsilon/2k_BT)]$, and $L^{(n)}_3=\int_{-\infty}^\infty \varepsilon^n d\varepsilon[1+3\tanh^4(\varepsilon/2k_BT)-4\tanh^2(\varepsilon/2k_BT)]$.
The calculation of these integrals gives
\begin{eqnarray}
L^{(n)}_1=
\left\{\begin{array}{ll}
-4k_BT& (n=0)\\
0 &(n=1)\\
-4\pi^2k_B^3T^3/3 &(n=2)
\end{array}\right.~,
\end{eqnarray}
\begin{eqnarray}
L^{(n)}_2 =
\left\{\begin{array}{ll}
0& (n=0)\\
-4k_B^2T^2 &(n=1)\\
0&(n=2)
\end{array}\right.~,
\end{eqnarray}
and
\begin{eqnarray}
L^{(n)}_3=
\left\{\begin{array}{ll}
0& (n=0)\\
0 &(n=1)\\
-16k_B^3T^3 &(n=2)
\end{array}\right.~.
\end{eqnarray}

Finally, we get
\begin{eqnarray}\label{I0}
 &&I^{(0)}_{\gamma\delta}=(\hbar\omega+\mu_\delta-\mu_\gamma)N(\hbar\omega+\mu_\delta-\mu_\gamma)~,\\\label{I1}
 &&I^{(1)}_{\gamma\delta}=\frac{(\hbar\omega+\mu_\delta)^2-\mu_\gamma^2}{2}N(\hbar\omega+\mu_\delta-\mu_\gamma)~,\\\label{I2}
&& I^{(2)}_{\gamma\delta}=\bigg[\frac{(\hbar\omega+\mu_\delta-\mu_\gamma)\pi^2k_B^2T^2}{3}\nonumber\\
 &&+\frac{(\hbar\omega+\mu_\delta)^3-\mu_\gamma^3}{3}\bigg]N(\hbar\omega+\mu_\delta-\mu_\gamma)~.
\end{eqnarray}

\subsection{Limit of weak transmission $\mathcal{T}\ll 1$}

In this subsection, we give the calculation of the expressions appearing in the first column of Table I.

\subsubsection{Electrical noise spectrum}

We calculate only $ \mathcal{S}^{00}_{LL}(\omega)$ since when $t$ and $\mathcal{T}$ are independent of energy, we have the relations
\begin{eqnarray*}
 \mathcal{S}^{00}_{LL}(\omega)&=&\mathcal{S}^{00}_{RR}(\omega)=-\mathcal{S}^{00}_{LR}(\omega)=-\mathcal{S}^{00}_{RL}(\omega)~.
\end{eqnarray*}
At weak $\mathcal{T}$, we have
\begin{eqnarray}
&&\mathcal{S}^{00}_{LL}(\omega)=\frac{e^{2}}{h}\sum_{\gamma\delta}\int_{-\infty}^\infty d\varepsilon \mathcal{M}_{LL}^{\gamma\delta}f^e_\gamma(\varepsilon)f^h_\delta(\varepsilon-\hbar\omega)\nonumber\\
&=&\frac{e^{2}\mathcal{T}}{h}\int_{-\infty}^\infty [f^e_L(\varepsilon)f^h_R(\varepsilon-\hbar\omega)
+f^e_R(\varepsilon)f^h_L(\varepsilon-\hbar\omega)]d\varepsilon\nonumber\\
&=&\frac{e^{2}\mathcal{T}}{h}\Big[ I^{(0)}_{LR}+ I^{(0)}_{RL}\Big]~,
\end{eqnarray}
which gives
\begin{eqnarray}
\mathcal{S}^{00}_{LL}(\omega)&=&\frac{e^{2}\mathcal{T}}{h}\big[(\hbar\omega-eV)N(\hbar\omega-eV)\nonumber\\
&&+(\hbar\omega+eV)N(\hbar\omega+eV)\big]~.
\end{eqnarray}
It reduces at equilibrium (zero-voltage) to
\begin{eqnarray}
\mathcal{S}^{00}_{LL}(\omega)&=&\frac{2e^{2}\mathcal{T}}{h}\hbar\omega N(\hbar\omega)~,
\end{eqnarray}
and at zero-temperature to 
\begin{eqnarray}
\mathcal{S}^{00}_{LL}(\omega)
&=&\frac{e^{2}\mathcal{T}}{h}\big[(eV-\hbar\omega)\Theta(eV-\hbar\omega)\nonumber\\
&&-(\hbar\omega+eV)\Theta(-eV-\hbar\omega)\big]~,
\end{eqnarray}
since we have $N(x)=-\Theta(-x)$ when $T\rightarrow 0$.

\subsubsection{Mixed noise spectrum}

We start from
\begin{eqnarray}
\mathcal{S}^{01}_{\alpha\beta}(\omega)&=&\frac{e}{h}\sum_{\gamma\delta}\int_{-\infty}^{\infty} d\varepsilon 
\big[(\varepsilon-\mu_\beta)\mathcal{M}_{\alpha\beta}^{\gamma\delta}\nonumber\\
&&-\hbar\omega\mathcal{N}_{\alpha\beta}^{\gamma\delta}\big]f^e_\gamma(\varepsilon)f^h_\delta(\varepsilon-\hbar\omega)~.
\end{eqnarray}
We calculate only $ \mathcal{S}^{01}_{LL}(\omega)$ since at weak $\mathcal{T}$, we have:
\begin{eqnarray*}
 \mathcal{S}^{01}_{LL}(\omega)&=&-\mathcal{S}^{01}_{RR}(\omega)=\mathcal{S}^{01}_{LR}(\omega)=-\mathcal{S}^{01}_{RL}(\omega)~.
\end{eqnarray*}
We have
\begin{eqnarray}
\mathcal{S}^{01}_{LL}(\omega)&=&\frac{e\mathcal{T}}{h}\int_{-\infty}^{\infty} d\varepsilon 
\big[(\varepsilon-\mu_L)[f^e_L(\varepsilon)f^h_R(\varepsilon-\hbar\omega)\nonumber\\
&&+f^e_R(\varepsilon)f^h_L(\varepsilon-\hbar\omega)]-\hbar\omega f^e_R(\varepsilon)f^h_L(\varepsilon-\hbar\omega)\big]\nonumber\\
&=&\frac{e\mathcal{T}}{h}\Big[I^{(1)}_{LR}-\mu_L I^{(0)}_{LR}+I^{(1)}_{RL}-(\mu_L+\hbar\omega)I^{(0)}_{RL}\Big]~,\nonumber\\
\end{eqnarray}
which gives
\begin{eqnarray}
\mathcal{S}^{01}_{LL}(\omega)&=&\frac{e\mathcal{T}}{2h}\big[(\hbar\omega-eV)^2N(\hbar\omega-eV)\nonumber\\
&&-(\hbar\omega+eV)^2N(\hbar\omega+eV)\big]~.
\end{eqnarray}
It reduces to $\mathcal{S}^{01}_{LL}(\omega)=0$ at  equilibrium (zero-voltage)
and to 
\begin{eqnarray}
\mathcal{S}^{01}_{LL}(\omega)&=&\frac{e\mathcal{T}}{2h}\big[-(\hbar\omega-eV)^2\Theta(eV-\hbar\omega)\nonumber\\
&&+(\hbar\omega+eV)^2\Theta(-eV-\hbar\omega)\big]~,
\end{eqnarray}
at zero-temperature.

\subsubsection{Heat noise spectrum}

We start from
\begin{eqnarray}
\mathcal{S}^{11}_{\alpha\beta}(\omega)&=&\frac{1}{h}\sum_{\gamma\delta}\int_{-\infty}^{\infty} d\varepsilon 
\big[(\varepsilon-\mu_\alpha)(\varepsilon-\mu_\beta)\mathcal{M}_{\alpha\beta}^{\gamma\delta}\nonumber\\
&&-\hbar\omega(\varepsilon-\mu_\alpha)\mathcal{N}_{\alpha\beta}^{\gamma\delta}-\hbar\omega(\varepsilon-\mu_\beta)(\mathcal{N}_{\beta\alpha}^{\gamma\delta})^*\nonumber\\
&&+\hbar^2\omega^2\mathcal{O}_{\alpha\beta}^{\gamma\delta}\big]f^e_\gamma(\varepsilon)f^h_\delta(\varepsilon-\hbar\omega)~,
\end{eqnarray}
which gives for $\mathcal{S}^{11}_{LL}(\omega)$
\begin{eqnarray}
&&\mathcal{S}^{11}_{LL}(\omega)=\frac{\mathcal{T}}{h}\int_{-\infty}^{\infty} d\varepsilon 
\big[(\varepsilon-\mu_L)^2[f^e_L(\varepsilon)f^h_R(\varepsilon-\hbar\omega)\nonumber\\
&&+f^e_R(\varepsilon)f^h_L(\varepsilon-\hbar\omega)]-2\hbar\omega(\varepsilon-\mu_L)f^e_R(\varepsilon)f^h_L(\varepsilon-\hbar\omega)\nonumber\\
&&+\hbar^2\omega^2[f^e_L(\varepsilon)f^h_L(\varepsilon-\hbar\omega)+f^e_R(\varepsilon)f^h_L(\varepsilon-\hbar\omega)]\big]\nonumber\\
&&=\frac{\mathcal{T}}{h}
\bigg[I^{(2)}_{LR}+I^{(2)}_{RL}-2\mu_L(I^{(1)}_{LR}+I^{(1)}_{RL})+\mu_L^2(I^{(0)}_{LR}\nonumber\\
&&+I^{(0)}_{RL})-2\hbar\omega(I^{(1)}_{RL}-\mu_LI^{(0)}_{RL})
+\hbar^2\omega^2(I^{(0)}_{LL}+I^{(0)}_{RL})\bigg]~.\nonumber\\
\end{eqnarray}
We report the expressions of the integrals given by Eqs.~(\ref{I0})-(\ref{I2}) and factorize the various contributions. It gives
\begin{eqnarray}
\mathcal{S}^{11}_{LL}(\omega)&=&\frac{\mathcal{T}}{h}\bigg[(\hbar\omega)^3N(\hbar\omega)\nonumber\\
&&+\frac{\pi^2k_B^2T^2}{3}\Big[(\hbar\omega-eV)N(\hbar\omega-eV)\nonumber\\
&&+(\hbar\omega+eV)N(\hbar\omega+eV)\Big]\nonumber\\
&&+\frac{(\hbar\omega-eV)^3}{3}N(\hbar\omega-eV)\nonumber\\
&&+\frac{(\hbar\omega+eV)^3}{3}N(\hbar\omega+eV)\bigg]~,
\end{eqnarray}
which reduces at  equilibrium (zero-voltage) to 
\begin{eqnarray}
\mathcal{S}^{11}_{LL}(\omega)&=&\frac{\mathcal{T}}{h}\bigg[\frac{5}{3}(\hbar\omega)^3+\frac{2\pi^2k_B^2T^2}{3}\hbar\omega \bigg]N(\hbar\omega)~,\nonumber\\
\end{eqnarray}
and at zero-temperature to 
\begin{eqnarray}
\mathcal{S}^{11}_{LL}(\omega)&=&\frac{\mathcal{T}}{h}\bigg[-(\hbar\omega)^3\Theta(-\hbar\omega)+\nonumber\\
&&\frac{(eV-\hbar\omega)^3}{3}\Theta(eV-\hbar\omega)\nonumber\\
&&-\frac{(\hbar\omega+eV)^3}{3}\Theta(-eV-\hbar\omega)\bigg]~,
\end{eqnarray}
since we have $N(x)=-\Theta(-x)$ when $T\rightarrow 0$. Note that $\mathcal{S}^{11}_{RR}(\omega)$ is obtained from the expression of $\mathcal{S}^{11}_{LL}(\omega)$ by inverting the voltage $V\rightarrow -V$, as a consequence we have $\mathcal{S}^{11}_{RR}(\omega)=\mathcal{S}^{11}_{LL}(\omega)$. We now calculate
\begin{eqnarray}
\mathcal{S}^{11}_{LR}(\omega)&=&\frac{\mathcal{T}}{h}\int_{-\infty}^{\infty} d\varepsilon 
\Big[-(\varepsilon-\mu_L)(\varepsilon-\mu_R)\nonumber\\
&&\times[f^e_L(\varepsilon)f^h_R(\varepsilon-\hbar\omega)+f^e_R(\varepsilon)f^h_L(\varepsilon-\hbar\omega)]\nonumber\\
&&+\hbar\omega(\varepsilon-\mu_L)f^e_L(\varepsilon)f^h_R(\varepsilon-\hbar\omega)\nonumber\\
&&
+\hbar\omega(\varepsilon-\mu_R)f^e_R(\varepsilon)f^h_L(\varepsilon-\hbar\omega)\Big]\nonumber\\
&=&\frac{\mathcal{T}}{h}\Big[-I^{(2)}_{LR}-I^{(2)}_{RL}+(\mu_L+\mu_R)\big[I^{(1)}_{LR}+I^{(1)}_{RL}\big]\nonumber\\
&&-\mu_L\mu_R\big[I^{(0)}_{LR}+I^{(0)}_{RL}\big]+\hbar\omega\big[I^{(1)}_{LR}+I^{(1)}_{RL}\big]\nonumber\\
&&-\hbar\omega\mu_LI^{(0)}_{LR} -\hbar\omega\mu_RI^{(0)}_{RL} \Big]~.
\end{eqnarray}
We report the expressions of the integrals given by Eqs.~(\ref{I0})-(\ref{I2}) and factorize the various contributions. It gives
\begin{eqnarray}
\mathcal{S}^{11}_{LR}(\omega)&=&\frac{\mathcal{T}}{h}\bigg[-\frac{\pi^2k_B^2T^2}{3}\Big[(\hbar\omega-eV)N(\hbar\omega-eV)\nonumber\\
&&+(\hbar\omega+eV)N(\hbar\omega+eV)\Big]\nonumber\\
&&+\frac{(\hbar\omega-eV)^3}{6}N(\hbar\omega-eV)\nonumber\\
&&+\frac{(\hbar\omega+eV)^3}{6}N(\hbar\omega+eV)\bigg]~,
\end{eqnarray}
which reduces at  equilibrium (zero-voltage) to 
\begin{eqnarray}
\mathcal{S}^{11}_{LR}(\omega)&=&\frac{\mathcal{T}}{h}\bigg[\frac{1}{3}(\hbar\omega)^3-\frac{2\pi^2k_B^2T^2}{3}\hbar\omega \bigg]N(\hbar\omega)~,\nonumber\\
\end{eqnarray}
and at zero-temperature to 
\begin{eqnarray}
\mathcal{S}^{11}_{LR}(\omega)&=&\frac{\mathcal{T}}{6h}\bigg[(eV-\hbar\omega)^3\Theta(eV-\hbar\omega)\nonumber\\
&&-(\hbar\omega+eV)^3\Theta(-eV-\hbar\omega)\bigg]~,
\end{eqnarray}
since we have $N(x)=-\Theta(-x)$ when $T\rightarrow 0$. Note that $\mathcal{S}^{11}_{RL}(\omega)$ is obtained from the expression of $\mathcal{S}^{11}_{LR}(\omega)$ by inverting the voltage $V\rightarrow -V$, as a consequence we have $\mathcal{S}^{11}_{RL}(\omega)=\mathcal{S}^{11}_{LR}(\omega)$.\\

\subsection{Limit of perfect transmission $\mathcal{T}=1$}

In this subsection, we give the calculation of the expressions appearing in the second column of Table I.

\subsubsection{Electrical noise spectrum}

For $\mathcal{T}=1$, we have
\begin{eqnarray}
 \mathcal{S}^{00}_{\alpha\beta}(\omega)&=&\frac{e^{2}}{h}(2\delta_{\alpha\beta}-1)\int_{-\infty}^\infty d\varepsilon
[f^e_L(\varepsilon)f^h_L(\varepsilon-\hbar\omega)\nonumber\\
&&+f^e_R(\varepsilon)f^h_R(\varepsilon-\hbar\omega)]\nonumber\\
&=&\frac{e^{2}}{h}(2\delta_{\alpha\beta}-1)[I^{(0)}_{LL}+I^{(0)}_{RR}]\nonumber\\
&=&\frac{e^{2}}{h}(2\delta_{\alpha\beta}-1)2\hbar\omega N(\hbar\omega)~.
\end{eqnarray}

\subsubsection{Mixed noise spectrum}

For $\mathcal{T}=1$, we have
\begin{eqnarray}
 &&\mathcal{S}^{01}_{LL}(\omega)=\frac{e}{h}\int_{-\infty}^\infty d\varepsilon
\left(\varepsilon-\mu_L-\frac{\hbar\omega}{2}\right)\nonumber\\
&&\times\Big[f^e_L(\varepsilon)f^h_L(\varepsilon-\hbar\omega)+f^e_R(\varepsilon)f^h_R(\varepsilon-\hbar\omega)\Big]\nonumber\\
&&=\frac{e}{h}\left[I^{(1)}_{LL}+I^{(1)}_{RR}-\left(\mu_L+\frac{\hbar\omega}{2}\right)[I^{(0)}_{LL}+I^{(0)}_{RR}]\right]~.\nonumber\\
\end{eqnarray}
After simplification, it leads to 
\begin{eqnarray}
\mathcal{S}^{01}_{LL}(\omega)=-\frac{e}{h}\hbar\omega eV N(\hbar\omega)~.
\end{eqnarray}
A similar calculation leads to $\mathcal{S}^{01}_{RR}(\omega)=\frac{e}{h}\hbar\omega eV N(\hbar\omega)$. Moreover, we have $ \mathcal{S}^{01}_{LR}(\omega)=-\mathcal{S}^{01}_{RR}(\omega)$, and $\mathcal{S}^{01}_{RL}(\omega)=-\mathcal{S}^{01}_{LL}(\omega)$.

\subsubsection{Heat noise spectrum}

For $\mathcal{T}=1$, we have
\begin{eqnarray}
 &&\mathcal{S}^{11}_{LL}(\omega)=\frac{1}{h}\int_{-\infty}^\infty d\varepsilon\bigg([(\varepsilon-\mu_L)^2-\hbar\omega(\varepsilon-\mu_L)]\nonumber\\
&&\times[f^e_L(\varepsilon)f^h_L(\varepsilon-\hbar\omega)+f^e_R(\varepsilon)f^h_R(\varepsilon-\hbar\omega)]\nonumber\\
&&
+\frac{\hbar^2\omega^2}{4}\sum_{\gamma\delta}f^e_\gamma(\varepsilon)f^h_\delta(\varepsilon-\hbar\omega)\bigg)\nonumber\\
&&=\frac{1}{h}\bigg[I^{(2)}_{LL}+I^{(2)}_{RR}-(2\mu_L+\hbar\omega)[I^{(1)}_{LL}+I^{(1)}_{RR}]\nonumber\\
&&+\mu_L(\mu_L+\hbar\omega)[I^{(0)}_{LL}+I^{(0)}_{RR}]+\frac{\hbar^2\omega^2}{4}\sum_{\gamma\delta}I^{(0)}_{\alpha\beta}\bigg]~.\nonumber\\
\end{eqnarray}
It gives
\begin{eqnarray}
&&\mathcal{S}^{11}_{LL}(\omega)=\frac{1}{h}\bigg[\left(\frac{2\hbar\omega\pi^2k_B^2T^2}{3}+\frac{\hbar^3\omega^3}{6}+\hbar\omega e^2V^2\right)N(\hbar\omega)\nonumber\\
&&+\frac{\hbar^2\omega^2}{4}\sum_{\pm}(\hbar\omega\pm  eV)N(\hbar\omega\pm  eV)\bigg]~.
\end{eqnarray}
A similar calculation leads to
\begin{eqnarray}
&&\mathcal{S}^{11}_{LR}(\omega)
=-\frac{1}{h}\bigg[\left(\frac{2\hbar\omega\pi^2k_B^2T^2}{3}+\frac{\hbar^3\omega^3}{6}\right)N(\hbar\omega)\nonumber\\
&&+\frac{\hbar^2\omega^2}{4}\sum_{\pm}(\hbar\omega\pm  eV)N(\hbar\omega\pm  eV)\bigg]~.
\end{eqnarray}
In addition, we have $\mathcal{S}^{11}_{RR}(\omega)=\mathcal{S}^{11}_{LL}(\omega)$ and $\mathcal{S}^{11}_{RL}(\omega)=\mathcal{S}^{11}_{LR}(\omega)$.



\begin{thebibliography}{99}
\bibitem{note0}For an energy independent transmission amplitude, we have $\mathcal{S}^{00}_{\alpha\alpha}(\omega)=-\mathcal{S}^{00}_{\alpha\bar\alpha}(\omega)$, and $\mathcal{S}^{01}_{L\alpha}(\omega)=-\mathcal{S}^{01}_{R\alpha}(\omega)$.
\bibitem{note1}Note that at zero-temperatures, for positive voltage and in weak transmission limit, the expressions for the noises read as
$\mathcal{S}^{pq}_{LR}(\omega)=e^{2-p-q}\mathcal{T}h^{-1}[(-1)^{1+p}|eV-\hbar\omega|^{1+p+q}\Theta(eV-\hbar\omega)
+(-1)^{1+q}|eV+\hbar\omega|^{1+p+q}\Theta(-eV-\hbar\omega)]/(1+p+q)!$,
 with $\mathcal{S}^{00}_{LL}(\omega)=-\mathcal{S}^{00}_{LR}(\omega)$, $\mathcal{S}^{01}_{LL}(\omega)=\mathcal{S}^{01}_{LR}(\omega)$, and
$\mathcal{S}^{11}_{LL}(\omega)=2\mathcal{S}^{11}_{LR}(\omega)+\mathcal{T}h^{-1}|\hbar\omega|^3\Theta(-\omega)$, where $\Theta$ is the Heaviside function.
\bibitem{noteHT}All the asymptotic values of noises given in the large temperature limit are calculated from Eqs.~(\ref{exp_noise})-(\ref{C}) taking $f_{\alpha,M}^{e,h}(\varepsilon)=1/2$. We have obtained: $\mathcal{S}^{00}_{\alpha\beta}(\omega)=e^2\pi\Gamma\delta_{\alpha\beta}/h$, $\mathcal{S}^{01}_{\alpha\beta}(\omega)=-e^2V\pi\Gamma\delta_{\alpha\beta}/h$, $\mathcal{S}^{11}_{\alpha\alpha}(\omega)=\hbar^2\omega^2\pi\Gamma/h$, and have checked numerically that $\mathcal{S}^{11}_{\alpha\bar\alpha}(\omega)$ varies linearly with temperature.
\end{thebibliography}
\end{document}